\documentclass[journal]{IEEEtran}
\usepackage{amssymb}
\usepackage{amsbsy}
\usepackage{amsmath}
\usepackage{amsthm} 
\usepackage{setspace}
\usepackage{epsfig}
\usepackage{cite}
\usepackage{graphics}
\usepackage{epstopdf}
\usepackage{mathrsfs}
\usepackage[above]{placeins}
\usepackage{algorithm}
\usepackage{algpseudocode}
\usepackage{multirow}
\usepackage{algorithmicx}
\usepackage{tikz}
\usepackage{stfloats}

\begin{document}
%


\title{\huge{Markov chain Monte Carlo Methods For Lattice Gaussian Sampling: Convergence Analysis and Enhancement}}



\author{Zheng~Wang,~\IEEEmembership{Member, IEEE}
\thanks{This work was presented in part at IEEE Information Theory Workshop (ITW), Kaohsiung, Taiwan, November, 2017 and IEEE International Symposium on Information Theory (ISIT), Honolulu, USA, June, 2014.

This work was supported in part by the National Natural Science Foundation of China under Grant 61801216, in part by the Natural Science Foundation of Jiangsu Province
under Grant SBK2018042902.

Z. Wang is with College of Electronic and Information Engineering, Nanjing University of Aeronautics and Astronautics (NUAA), Nanjing, China (e-mail: z.wang@ieee.org).

}}

\maketitle


\begin{abstract}
Sampling from lattice Gaussian distribution has emerged as an important problem in coding, decoding and cryptography. In this paper, the classic Gibbs algorithm from Markov chain Monte Carlo (MCMC) methods is demonstrated to be geometrically ergodic for lattice Gaussian sampling, which means the Markov chain arising from it converges exponentially fast to the stationary distribution. Meanwhile, the exponential convergence rate of Markov chain is also derived through the spectral radius of forward operator. Then, a comprehensive analysis regarding to the convergence rate is carried out and two sampling schemes are proposed to further enhance the convergence performance. The first one, referred to as Metropolis-within-Gibbs (MWG) algorithm, improves the convergence by refining the state space of the univariate sampling. On the other hand, the blocked strategy of Gibbs algorithm, which performs the sampling over multivariate at each Markov move, is also shown to yield a better convergence rate than the traditional univariate sampling. In order to perform blocked sampling efficiently, Gibbs-Klein (GK) algorithm is proposed, which samples block by block using Klein’s algorithm. Furthermore, the validity of GK algorithm is demonstrated by showing its ergodicity. Simulation results based on MIMO detections are presented to confirm the convergence gain brought by the proposed Gibbs sampling schemes.
\end{abstract}

\IEEEpeerreviewmaketitle

\vspace{0.5em}

\IEEEpeerreviewmaketitle

\section{Introduction}
Nowadays, lattice Gaussian distribution, which is supported over a multi-dimensional Euclidean lattice, has drawn a lot of attentions in various research fields. In mathematics, Banaszczyk first applied it to prove the transference theorems for lattices \cite{Banaszczyk}. In coding, lattice Gaussian distribution was employed to obtain the full shaping gain for lattice coding \cite{Forney_89,Kschischang_Pasupathy,LiuLing2}, and to achieve the capacity of the Gaussian channel \cite{LB_13}. Meanwhile, it was also used to achieve information-theoretic security in Gaussian wiretap channel \cite{LiuLing1,LLBS_12,7360779}. Furthermore, lattice Gaussian distribution has been adapted to bidirectional relay network under the compute-and-forward strategy for the physical layer security \cite{7058433}. Additionally, it is also applied to realize the probabilistic shaping for optical communication systems \cite{ProbabilisticShapingOptical1,ProbabilisticShapingOptical2}.

In cryptography, lattice Gaussian distribution has already become a central tool in the construction of many primitives. Specifically, Micciancio and Regev used it to propose lattice-based cryptosystems based on the worst-case hardness assumptions \cite{MicciancioGaussian}. In \cite{CHKP10}, lattice Gaussian distribution is applied to create even more powerful cryptographic primitives, namely, hierarchical identity-based encryption and standard model signatures. In learning with errors (LWE) based encryption, sampling from lattice Gaussian distribution is highly demanded for the key generation \cite{LwelatticeGaussian}. Meanwhile, it also has underpinned the fully-homomorphic encryption for cloud computing \cite{GentrySW13}. Besides, there are various applications that require sampling over lattice Gaussian distribution as part of the ``on-line" computation, where the most notable one among them is the secure lattice-based digital signature on a constrain device \cite{latticeGaussianSignatures}.

On the other hand, algorithmically, lattice Gaussian distribution with a suitable variance allows lattice decoding to solve the shortest vector problem (SVP) and the closest vector problem (CVP) \cite{RegevSolvingtheShortestVectorProblem,RegevSolvingtheClosestVectorProblem}. Intuitively, the formulation of it comes from a conceptually simple fact that each lattice point in the discrete Gaussian distribution entails a sampling probability scaled by the Euclidean distance from the query point \cite{Klein}. The lattice points who are close to the center of the distribution naturally correspond to large sampling probabilities.
Therefore, the desired closest lattice point or shortest lattice vector would conceivably be obtained due to the largest sampling probability. To this end, sampling over lattice Gaussian distribution has been widely applied in multi-input multi-output (MIMO) communications for signal detection \cite{ZhengWangTIT17,DerandomizedJ,CongRandom}. Compared to the optimal sphere detection, it is not only much more efficient, but also can be realized flexibly to achieve the trade-off between decoding performance and complexity \cite{Shaoshi1}. In addition, such a sampling decoding strategy can be easily extended to signal processing as an useful signal estimator or detector \cite{Luo1,xia2,Wuqihui1,Zhuang1,Xiangmin1}.

%


Because of the central role of lattice Gaussian distribution playing in these fields, its sampling algorithms become an important computational problem. However, different from the case of continuous Gaussian density, sampling from the discrete lattice Gaussian distribution is by no means trivial even for a low-dimensional system. For this reason, Markov chain Monte Carlo (MCMC) methods were introduced as an alternative way for lattice Gaussian sampling, which attempts to sample from the target distribution by building a Markov chain \cite{ZhengWangTIT15,ZhengWangMCMCLatticeGaussian}. Typically, after a burn-in stage, which is normally measured by the \emph{mixing time} in total variance distance, the Markov chain will step into a stationary distribution, where samples from the target distribution can be successfully obtained thereafter. Therefore, in MCMC, the complexity of each Markov move is normally insignificant, whereas the required mixing time as well as the convergence rate are more critical.

Specifically, in \cite{ZhengWangMCMCLatticeGaussian}, Gibbs algorithm was introduced into lattice Gaussian sampling by showing its ergodicity, which employs conditional univariate sampling to build the Markov chain. Nevertheless, ergodicity only guarantees the convergence while the way of convergence (e.g., polynomial convergence, geometric convergence and so on) as well as the related convergence rate are unclear, resulting in an untractable Markov chain. In fact, compared to Gibbs algorithm for lattice Gaussian distribution, a better progress has been made with respect to Metropolis-Hastings (MH) algorithm, which is well known as another important sampling scheme in MCMC. For example, the Metropolis-Hastings-Klein (MHK) algorithm given in \cite{ZhengWangTIT15} is not only uniformly ergodic in tackling with lattice Gaussian sampling, but also enjoys an accessible convergence rate.

This paper was previously presented at conferences \cite{ZhengWangMCMCLatticeGaussian} and \cite{ITW2017} while the following extensions are given. On one hand, with respect to the geometric ergodicity of Gibbs algorithm for lattice Gaussian sampling \cite{ITW2017}, a prospective way for convergence diagnosis by means of
the spectral radius of the forward operator is offered. Inspired by it, convergence analysis
is carried out in this paper, where the corresponding enhancement scheme named as Metropolis-within-Gibbs (MWG) algorithm is proposed for univariate Gibbs sampling. More importantly, the superiority of MWG over Gibbs algorithm in terms of convergence rate is demonstrated, and further improvement can be realized by the parallel tempering technique. On the other hand, different from the previous work \cite{ZhengWangMCMCLatticeGaussian} which only concerns the efficient implementation for the blocked strategy of Gibbs algorithm regardless of the convergence behaviour, the blocked strategy by sampling over multivariate is demonstrated to enable a faster convergence rate than the univariate sampling. Moreover, the geometric ergodicity of the proposed Gibbs-Klein (GK) algorithm is also given, which removes the approximation errors by resorting to the rejection sampling. Hence, a whole framework of Gibbs-based algorithms for lattice Gaussian sampling is established. In addition, the convergence gain due to the proposed enhancement schemes are confirmed through the context of signal detection in MIMO systems.

The rest of this paper is organized as follows. Section II introduces the lattice Gaussian distribution and briefly reviews the basics of MCMC methods. In Section III, Gibbs algorithm is introduced to lattice Gaussian sampling, and its geometric ergodicity is demonstrated. In Section IV, the Metropolis-within-Gibbs algorithm is proposed to strengthen the convergence performance in terms of the univariate sampling, followed by the proof to show the convergence enhancement. In Section V, blocked strategy is adopted to Gibbs algorithm to achieve a better convergence rate. In order to realize efficient blocked sampling, Gibbs-Klein algorithm is proposed and the proof of its validity is also given. Simulations through MIMO systems are presented in Section VI to illustrate the convergence gain of these two proposed algorithms. At the end, Section VII concludes the paper.

\emph{Notation:} Matrices and column vectors are denoted by upper and lowercase boldface letters, and the transpose, inverse, pseudoinverse of a matrix $\mathbf{B}$ by $\mathbf{B}^T, \mathbf{B}^{-1},$ and $\mathbf{B}^{\dag}$, respectively. We use $\mathbf{b}_i$ for the $i$th column of the matrix $\mathbf{B}$, $\mathbf{\widehat{b}}_i$ for the $i$th
Gram-Schmidt vector of the matrix $\mathbf{B}$, $b_{i,j}$ for the entry in the $i$th row and $j$th column of the matrix $\mathbf{B}$. $\lceil x \rfloor$ denotes rounding to
the integer closest to $x$. If $x$ is a complex number, $\lceil x \rfloor$ rounds the real and imaginary parts separately. In addition, in this paper, the computational complexity is measured by the number of arithmetic operations (additions, multiplications, comparisons, etc.). We use the standard \emph{small omega} notation $\omega(\cdot)$, i.e., $f(n) = \omega(g(n))$ if for any $k > 0$, the inequality $|f(n)|>k\cdot|g(n)|$ holds for all sufficiently large $n$. Finally, $h\in L_0^2(\pi)$ and $L_0^2(\pi)$ denote the set of all mean zero and finite variance functions with respect to the target distribution $\pi$, i.e., $E_{\pi}[h(\mathbf{x})]=0$ and $\text{var}_{\pi}[h(\mathbf{x})]=v<\infty$.

\newtheorem{my1}{Lemma}
\newtheorem{my2}{Theorem}
\newtheorem{my3}{Definition}
\newtheorem{my4}{Proposition}
\newtheorem{my5}{Remark}
\newtheorem{my6}{Conjection}
\newtheorem{my7}{Corollary}

\begin{figure}[t]
\vspace{-2em}
\hspace{-1em}\includegraphics[width=4in,height=2.7in]{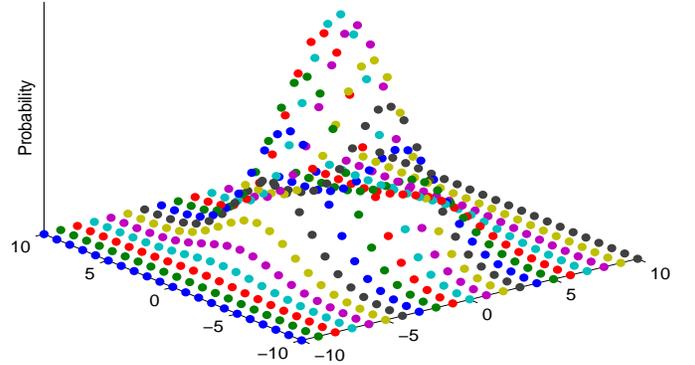}
\vspace{-2em}
  \caption{Illustration of a two-dimensional lattice Gaussian distribution.}
  \label{simulation x}
\end{figure}

\section{Preliminaries}
In this section, the background and mathematical tools needed to describe and analyze the Gibbs algorithm for lattice Gaussian sampling are introduced.

\subsection{Lattice Gaussian Distribution}
Let $\mathbf{B}=[\mathbf{b}_1,\ldots,\mathbf{b}_n]\subset \mathbb{R}^n$ consist of $n$ linearly independent vectors. The $n$-dimensional lattice $\Lambda$ generated by $\mathbf{B}$ is defined by
\begin{equation}
\Lambda=\mathcal{L}(\mathbf{B})=\{\mathbf{Bx}: \mathbf{x}\in \mathbb{Z}^n\},
\end{equation}
where the full rank matrix $\mathbf{B}\in\mathbb{R}^{n\times n}$ is called the lattice basis. The \emph{Gaussian function} centered at $\mathbf{c}\in \mathbb{R}^n$ with standard deviation $\sigma>0$ is defined as
\begin{equation}
\rho_{\sigma, \mathbf{c}}(\mathbf{z})=e^{-\frac{\|\mathbf{z}-\mathbf{c}\|^2}{2\sigma^2}},
\end{equation}
for all $\mathbf{z}\in\mathbb{R}^n$. When $\mathbf{c}$ or $\sigma$ are not specified, they are assumed to be $\mathbf{0}$ and $1$ respectively. Then, the \emph{discrete Gaussian distribution} over $\Lambda$ is defined as
\begin{equation}
D_{\Lambda,\sigma,\mathbf{c}}(\mathbf{x})=\frac{\rho_{\sigma, \mathbf{c}}(\mathbf{Bx})}{\rho_{\sigma, \mathbf{c}}(\Lambda)}=\frac{e^{-\frac{1}{2\sigma^2}\parallel \mathbf{Bx}-\mathbf{c} \parallel^2}}{\sum_{\mathbf{x} \in \mathbb{Z}^n}e^{-\frac{1}{2\sigma^2}\parallel \mathbf{Bx}-\mathbf{c} \parallel^2}}
\label{lattice gaussian distribution}
\end{equation}
for all $\mathbf{x}\in \mathbb{Z}^n$, where $\rho_{\sigma, \mathbf{c}}(\Lambda)\triangleq \sum_{\mathbf{\mathbf{Bx}}\in\Lambda}\rho_{\sigma, \mathbf{c}}(\mathbf{Bx})$ is just a scaling to obtain a probability distribution and $\sigma>0$ represents the standard deviation.



Note that this definition differs slightly from the one in \cite{MicciancioGaussian}, where $\sigma$ is scaled by a constant factor $\sqrt{2\pi}$ (i.e., $s =\sqrt{2\pi}\sigma$). Fig. 1 illustrates the discrete Gaussian distribution over $\mathbb{Z}^2$. As can be seen clearly, it resembles a continuous Gaussian distribution, but is only defined over a lattice. In fact, discrete and continuous
Gaussian distributions share similar properties, if the \emph{flatness
factor} is small \cite{LLBS_12}.


\subsection{Klein's Algorithm}
At present, the default sampling algorithm for lattice Gaussian distribution is due to Klein, which is originally proposed for bounded-distance decoding (BDD) in lattices \cite{Klein}. As shown in Algorithm 1, the operation of Klein's algorithm has polynomial complexity $O(n^2)$ excluding QR decomposition (which may be done only once at the beginning). More precisely, by sequentially sampling from the 1-dimensional conditional Gaussian distribution $D_{\mathbb{Z},\sigma_i,\widetilde{x}_i}$ in a backward order from $x_n$ to $x_1$ (the forward order from $x_1$ to $x_n$ is fine as well), the Gaussian-like distribution arising from Klein's algorithm is given by
\begin{eqnarray}
P_{\text{Klein}}(\mathbf{x})&=&\prod^n_{i=1}D_{\mathbb{Z},\sigma_i,\widetilde{x}_i}(x_i)=\frac{\rho_{\sigma, \mathbf{c}}(\mathbf{Bx})}{\prod^n_{i=1}\rho_{\sigma_i, \widetilde{x}_i}(\mathbb{Z})}\notag\\
&=&\frac{e^{-\frac{1}{2\sigma^2}\parallel \mathbf{Bx}-\mathbf{c} \parallel^2}}{\prod^n_{i=1}\sum_{\widetilde{x}_i\in\mathbb{Z}}e^{-\frac{1}{2\sigma^2_i}\|x_i-\widetilde{x}_i\|^2}},
\label{klein distribution}
\end{eqnarray}
where $\widetilde{x}_i=\frac{c'_i-\sum^n_{j=i+1}r_{i,j}x_j}{r_{i,i}}$, $\sigma_i=\frac{\sigma}{|r_{i,i}|}=\frac{\sigma}{\|\mathbf{\widehat{b}}_i\|}$, $\mathbf{c'}=\mathbf{Q}^{\dag}\mathbf{c}$, $r_{i,j}$ denotes the element of the upper triangular matrix $\mathbf{R}$ from the QR decomposition $\mathbf{B}=\mathbf{QR}$ and $\mathbf{\widehat{b}}_i$'s are the Gram-Schmidt vectors of $\mathbf{B}$ with $\|\mathbf{\widehat{b}}_i\|=|r_{i,i}|$.

In \cite{Trapdoor}, it has been demonstrated that $P_{\text{Klein}}(\mathbf{x})$ is close to $D_{\Lambda,\sigma,\mathbf{c}}(\mathbf{x})$ within a negligible statistical distance if
\begin{equation}
\sigma \geq \omega(\sqrt{\log n})\cdot\max_{1\leq i \leq n}\|\mathbf{\widehat{b}}_i\|.
\end{equation}
However, even with the help of lattice reduction\footnote{It is well known that lattice reduction such as the Lenstra-Lenstra-Lov\'{a}sz (LLL) algorithm is able to significantly improve $\min_i\|\widehat{\mathbf{b}}_i\|$ while reducing $\max_i\|\widehat{\mathbf{b}}_i\|$ at the same time \cite{LLLoriginal,Shanxiang1}.}, the requirement of standard deviation $\omega(\sqrt{\log n})\cdot\max_{1\leq i \leq n}\|\mathbf{\widehat{b}}_i\|$ is too large to be useful in practice, rendering Klein's algorithm inapplicable to many cases of interest.

\renewcommand{\algorithmicrequire}{\textbf{Input:}}  
\renewcommand{\algorithmicensure}{\textbf{Output:}} 

\begin{algorithm}[t]
\caption{Klein's Algorithm}
\begin{algorithmic}[1]
\Require
$\mathbf{B}, \sigma, \mathbf{c}$
\Ensure
$\mathbf{Bx}\in\Lambda$
\State let $\mathbf{B}=\mathbf{QR}$ and $\mathbf{c'}=\mathbf{Q}^{T}\mathbf{c}$
\For {$i=n$,\ \ldots,\ 1}
\State let $\sigma_i=\frac{\sigma}{|r_{i,i}|}$ and $\widetilde{x}_i=\frac{c'_i-\sum^n_{j=i+1}r_{i,j}x_j}{r_{i,i}}$
\State sample $x_i$ from $D_{\mathbb{Z},\sigma_i,\widetilde{x}_i}$
\EndFor
\State return $\mathbf{Bx}$
\end{algorithmic}
\end{algorithm}

\subsection{MCMC Methods}
As for lattice Gaussian sampling in the range $\sigma < \omega(\sqrt{\log n})\cdot\max_{1\leq i \leq n}\|\mathbf{\widehat{b}}_i\|$, MCMC methods have become an important feasible way, where the lattice Gaussian distribution $D_{\Lambda,\sigma,\mathbf{c}}$ is viewed as a complex target distribution lacking direct sampling methods. By establishing a Markov chain that randomly generates the next state, MCMC is capable of sampling from the target distribution of interest, thereby removing the restriction on $\sigma$.

As an important parameter which measures the time required by a
Markov chain to get close to its stationary distribution, the \emph{mixing time} is defined as \cite{mixingtimemarkovchain}
\begin{equation}
t_{\text{mix}}(\epsilon)=\text{min}\{t:\max\|P^t(\mathbf{x}, \cdot)-\pi(\cdot)\|_{TV}\leq \epsilon\},
\label{mixing time}
\end{equation}
where $\|\cdot\|_{TV}$ represents the total variation distance\footnote{Other measures of distance also exist, see \cite{RosenthalOPDAMCMC} for more details.}.
Thanks to the celebrated \emph{coupling technique}, for any Markov chain with finite state space $\Omega$, exponentially fast convergence can be demonstrated if the underlying Markov chain is irreducible and aperiodic with an invariant distribution $\pi$ \cite{mixingtimemarkovchain}. Nevertheless, in the case of lattice Gaussian sampling, the countably infinite state space $\mathbf{x}\in \mathbb{Z}^n$ naturally imposes a challenge. For this reason, we perform the convergence analysis from the beginning --- ergodicity \cite{mixingtimemarkovchain}.

\begin{my3}
Let $\mathbf{P}$ be an irreducible and aperiodic transition matrix for a Markov chain. If the chain is reversible (hence positive recurrent), then it is ergodic, namely, there is a unique probability distribution $\pi$ on $\Omega$ and for all $\mathbf{x}\in\Omega$,
\begin{equation}
\underset{t\rightarrow\infty}{{\lim}}\|P^t(\mathbf{x},\cdot)-\pi\|_{TV}=0,
\end{equation}
{where $P^t(\mathbf{x}; \cdot)$ denotes a row of the transition matrix $\mathbf{P}$ for $t$ Markov moves.}
\end{my3}

Note that the state space $\Omega$ in the below definition can be countably infinite. Unless stated otherwise, the state space of the Markov chain we are concerned with throughout the context is the countably infinite $\Omega=\mathbb{Z}^n$. Although \emph{ergodicity} implies asymptotic convergence to stationarity, it does not entail the way of convergence, resulting in an intractable Markov chain \cite{LukeTierney1}. Among kinds of ergodicity in literature \cite{Meynbook}, \emph{geometric ergodicity} which converges exponentially is defined as:

\begin{my3}
A Markov chain with stationary distribution $\pi$ is geometrically ergodic if there exists $0<\varrho<1$ and $M(\mathbf{x})<\infty$ such that for all $\mathbf{x}$
\begin{equation}
\|P^t(\mathbf{x}, \cdot)-\pi(\cdot)\|_{TV}\leq M(\mathbf{x})\varrho^t,
\label{geo-ergodic}
\end{equation}
where $M(\mathbf{x})$ is parameterized by the initial state $\mathbf{x}$.
\end{my3}
Clearly, parameter $\varrho$ is the convergence rate of the Markov chain. Compared to geometric ergodicity,
\emph{uniform ergodicity} also converges exponentially but its convergence does not depend on the initial state $\mathbf{x}$, leading to a constant $M$ \cite{Meynbook}.

\section{Gibbs Algorithm for Lattice Gaussian Sampling}
In this section, Gibbs algorithm is introduced to lattice Gaussian sampling, which establishes the Markov chain through univariate sampling.

\subsection{Ergodicity}

\renewcommand{\algorithmicrequire}{\textbf{Input:}}  
\renewcommand{\algorithmicensure}{\textbf{Output:}} 

The origin of Gibbs algorithm can be traced back to the celebrated work of Geman in 1984 \cite{GibbsOriginal}. As a foremost sampling scheme in MCMC, Gibbs algorithm is widely applied in various research fields of mathematics, statistics and physics for obtaining a sequence of observations which are approximated from a specified multivariate probability distribution, when direct sampling is difficult. Specifically, it tries to tackles with the sampling from a complicated joint distribution through conditional sampling over its marginal distributions.

Typically, as for lattice Gaussian sampling by Gibbs algorithm, each coordinate of $\mathbf{x}$ is sampled from the following 1-dimensional conditional distribution
\begin{equation}
P_i(x_i|\mathbf{x}_{[-i]})\hspace{-.2em}=\hspace{-.2em}D_{\Lambda,\sigma,\mathbf{c}}(x_i|\mathbf{x}_{[-i]})\hspace{-.2em}=\hspace{-.2em}\frac{e^{-\frac{1}{2\sigma^2}\parallel \mathbf{Bx}-\mathbf{c} \parallel^2}}{\sum_{x_i\in \mathbb{Z}}e^{-\frac{1}{2\sigma^2}\parallel \mathbf{Bx}-\mathbf{c} \parallel^2}}
\label{x7}
\end{equation}
with $\sigma>0$. Here $1\leq i\leq n$ denotes the coordinate index of $\mathbf{x}$, $\mathbf{x}_{[-i]}\triangleq[x_1,\ldots,x_{i-1},x_{i+1},\ldots,x_{n}]^T$. During this univariate sampling, the other $n-1$ variables contained in $\mathbf{x}_{[-i]}$ are leaving unchanged. By repeating such a procedure with a certain scan scheme, a Markov chain $\{\mathbf{X}^0, \mathbf{X}^1, \ldots\}$ is established.

In Gibbs algorithm, there are various scan schemes to proceed the component updating. Among them, random scan is the basic one. Typically, under random scan, the coordinate index $i$ is randomly chosen from a set of selection probabilities $[\beta_1, \ldots, \beta_n]$, where $\sum_{i=1}^n\beta_i=1$ and $\beta_i>0$. The extension to other scan strategies is possible. Without loss of generality, the random scan scheme is considered for Gibbs algorithm throughput the context and flexible implementation based on it can be easily carried out in practice. Therefore, the transition probability $P(\mathbf{X}^t,\mathbf{X}^{t+1})$ of Gibbs algorithm for lattice Gaussian sampling is
\begin{equation}
P(\mathbf{X}^t=\mathbf{x}, \mathbf{X}^{t+1}=\mathbf{y})=P_{i}(x_i|\mathbf{x}_{[-i]}),
\label{eq:transition}
\end{equation}
where random variable $i$ follows from the distribution $[\beta_1, \ldots, \beta_n]$.
Clearly, every two adjacent states $\mathbf{X}^t=\mathbf{x}=[x^t_1,\ldots,x^t_i,\ldots,x^t_{n}]^T$ and $\mathbf{X}^{t+1}=\mathbf{y}=[x^t_1,\ldots,x^{t+1}_i,\ldots,x^t_{n}]^T$ differ from each other by only one coordinate $x_i$.

For a given standard deviation $\sigma>0$ and full rank lattice basis $\mathbf{B}$, it is easy to verify that each random variable $x_i$ is sampled with variance
\begin{equation}
\text{var}[x_i|\mathbf{x}_{[-i]}]=\kappa_i>0.
\label{correlationsetup}
\end{equation}
Therefore, all the sampling candidates of $x_i$ are possible to be sampled theoretically, indicating an irreducible chain. In principle, the irreducible property prevents the random variables to be totally dependent. To summarize, Algorithm 2 illustrates the operation of Gibbs algorithm for lattice Gaussian sampling. The initial Markov state $\mathbf{X}^0$ can be chosen from $\mathbb{Z}^n$ arbitrarily or from the output of a suboptimal algorithm, while $t_{\text{mix}}(\epsilon)$ denotes the mixing time that the Markov requires to step into the stationary distribution.

With the transition probabilities (\ref{eq:transition}), we may form the infinite transition matrix $\mathbf{P}$, whose $(i,j)$-th entry $P(\mathbf{x}_i; \mathbf{x}_j)$ represents the probability of transferring to state $\mathbf{x}_j$ from the previous state $\mathbf{x}_i$. Then, from Definition 1, besides irreducible property, it is also easy to verify that the underlying Markov chain is reversible and aperiodic so as to the following Theorem about ergodicity, where the proof is omitted due to simplicity.
\begin{my2}
Given the invariant distribution $D_{\Lambda,\sigma,\mathbf{c}}$, the Markov chain induced by the Gibbs algorithm is ergodic as
\end{my2}
\vspace{-1em}
\begin{equation}
\underset{t\rightarrow\infty}{\text{lim}}\|P^t(\mathbf{x}, \cdot)-D_{\Lambda,\sigma,\mathbf{c}}\|_{TV}=0,
\end{equation}
\emph{for all states $\mathbf{x}\in\mathbb{Z}^n$.}


\begin{algorithm}[t]
\caption{Gibbs Algorithm for Lattice Gaussian Sampling}
\begin{algorithmic}[1]
\Require
$\mathbf{B}, \sigma, \mathbf{c}, \mathbf{X}^0, \beta_i$'s, $t_{\text{mix}}(\epsilon)$
\Ensure
$\mathbf{x}\thicksim \pi,\ \pi\ \text{is within statistical distance of}\ \epsilon\ \text{to}\ D_{\Lambda,\sigma,\mathbf{c}}$
\For {$t=$1,2,\ \ldots}
\State let $\mathbf{x}$ denote the state of $\mathbf{X}^{t-1}$
\State randomly choose index $i$ by distribution $[\beta_1,\ldots,\beta_n]$
\State sample $x_i$ from $P_i(x_i|\mathbf{x}_{[-i]})$ shown in (\ref{x7})
\State update $\mathbf{x}$ with the sampled $x_i$ and let $\mathbf{X}^t=\mathbf{x}$
\If {$t\geq t_{\text{mix}}(\epsilon)$}
\State output the state of $\mathbf{X}^t$
\EndIf
\EndFor
\end{algorithmic}
\end{algorithm}

According to Theorem 1, if time permits to reach the stationary distribution, Gibbs algorithm will draw samples from $D_{\Lambda,\sigma,\mathbf{c}}$ no matter what value $\sigma>0$ is, which means the obstacle encountered by Klein's algorithm is overcome.

\subsection{Geometric Ergodicity}
With respect to convergence analysis, the notion of spectral gap $\gamma$ of the Markov chain is introduced to prove the geometric ergodicity\footnote{The geometric ergodicity of Markov chains can be also verified by other ways, i.e., \emph{drift condition} in \cite{RobertsGeneralstatespace,ZhengWangTIT15}.}. Specifically, according to the following Theorem from \cite{SpectralGapL2}, the geometric ergodicity can be verified by $\gamma>0$ for a reversible, irreducible and aperiodic Markov chain.

\begin{my2}[\hspace{-0.005em}\cite{SpectralGapL2}]
Given the invariant distribution $\pi$, a reversible, irreducible and aperiodic Markov chain with spectral gap $\gamma=1-\text{spec}(\mathbf{F})>0$ is geometric ergodicity
\end{my2}
\vspace{-1em}
\begin{equation}
\|P^t(\mathbf{x}, \cdot)-\pi(\cdot)\|_{TV}\leq M(\mathbf{x})(1-\gamma)^t,
\label{m1}
\end{equation}
\emph{where} $\text{spec}(\cdot)$ \emph{denotes the spectral radius and} $\mathbf{F}$ \emph{represents the forward operator of the Markov chain.}

Particularly, given the transition probability $P(\mathbf{X}^t, \mathbf{X}^{t+1})$, the forward operator $\mathbf{F}$ of the Markov chain is defined as \cite{LiuBook}
\begin{equation}
\mathbf{F}h(\mathbf{X}^t)\triangleq\hspace{-.8em}\sum_{\mathbf{X}^{t+1}\in\Omega}\hspace{-1em}h(\mathbf{X}^{t+1})P(\mathbf{X}^t, \mathbf{X}^{t+1})\hspace{-.2em}=\hspace{-.2em}E[h(\mathbf{X}^{t+1})|\mathbf{X}^t]
\label{xp}
\end{equation}
with induced operator norm
\begin{equation}
\|\mathbf{F}\|=\hspace{-1em}\underset{h\in L_0^2(\pi), \text{var}(h)=1}{\sup}\hspace{-1em}\|\mathbf{F}h\|.
\label{xp1}
\end{equation}
Here, $L^2(\pi)$ is the Hilbert space of square integrable functions with respect to $\pi$ so that $L_0^2(\pi)\triangleq\{h(\mathbf{x}):E[h(\mathbf{x})]=0, \text{var}[h(\mathbf{x})]<\infty\}$ denotes the subspace of $L^2(\pi)$ consisting of functions with zero mean relative to $\pi$. More precisely, for $h(\cdot),g(\cdot)\in L_0^2(\pi)$, the inner product defined by the space is
\begin{equation}
\langle h(\mathbf{x}),g(\mathbf{x})\rangle=E[h(\mathbf{x})g(\mathbf{x})]
\end{equation}
with variance
\begin{equation}
\text{var}_{\pi}[h(\mathbf{x})]=\langle h(\mathbf{x}), h(\mathbf{x})\rangle=\|h(\mathbf{x})\|^2.
\label{m3}
\end{equation}

Clearly, from Theorem 2, the convergence rate of the Markov chain is exactly characterized by the spectral radius of $\mathbf{F}$, i.e., $\varrho=\text{spec}(\mathbf{F})$. Based on it, we then arrive at the following Corollary to show the geometric ergodicity.

\begin{my7}
Given the invariant lattice Gaussian distribution $D_{\Lambda,\sigma,\mathbf{c}}$, the Markov chain induced by Gibbs algorithm is geometrically ergodic
\end{my7}
\vspace{-1em}
\begin{equation}
\|P^t(\mathbf{x}, \cdot)-D_{\Lambda,\sigma,\mathbf{c}}\|_{TV}\leq M(\mathbf{x})\varrho^t
\label{geo-ergodic}
\end{equation}
\emph{with convergence rate} $\varrho=\text{spec}(\mathbf{F})<1$.
\begin{proof}
First of all, $\text{spec}(\mathbf{F})$ is closely related with the norm of $\mathbf{F}$ as \cite{Fill1991,SpectralGapL2}
\begin{equation}
\text{spec}(\mathbf{F})=\underset{t\rightarrow\infty}{\lim}\|\mathbf{F}^t\|^{1/t}.
\label{x9}
\end{equation}

Meanwhile, reversibility corresponds to a self-adjoint operator $\mathbf{F}$ with \cite{LiuCovarianceScans}
\begin{equation}
\|\mathbf{F}^t\|=\|\mathbf{F}\|^t,
\end{equation}
then we have
\begin{equation}
\text{spec}(\mathbf{F})=\|\mathbf{F}\|.
\label{m2}
\end{equation}

Subsequently, according to (\ref{xp1}) and (\ref{m3}), the spectral radius of $\mathbf{F}$ is further expressed as
\begin{equation*}
\text{spec}(\mathbf{F})=\hspace{-1em}\underset{h\in L_0^2(\pi), \text{var}(h)=1}{\sup}\hspace{-1em}\|\mathbf{F}h\|\ \ \ \ \ \ \ \ \ \ \ \ \ \ \ \ \ \ \ \ \ \ \ \ \ \ \ \ \ \ \ \ \ \ \ \ \ \ \ \ \ \ \ \ \ \ \ \ \ \ \ \ \ \ \ \ \
\end{equation*}
\vspace{-1em}
\begin{eqnarray}
\hspace{-1.1em}&=&\hspace{-1.2em}\hspace{-1em}\underset{h\in L_0^2(\pi), \text{var}(h)=1}{\sup}\hspace{-1em}\{\text{var}[E[h(\mathbf{X}^{t+1})|\mathbf{X}^t]]\}^{\frac{1}{2}}\label{xnmb1}\\
\hspace{-1.1em}&\overset{(a)}{=}&\hspace{-1.2em}\hspace{-1em}\underset{h\in L_0^2(\pi), \text{var}(h)=1}{\sup}\hspace{-1em}\{\text{var}[\hspace{-.1em}h(\mathbf{X}^{t+1})\hspace{-.1em}]\hspace{-.2em}-\hspace{-.2em}E[\text{var}[h(\mathbf{X}^{t+1})|\mathbf{X}^t]]\}^{\frac{1}{2}}\notag\\
\hspace{-1.1em}&=&\hspace{-1.2em}\left[\underset{h\in L_0^2(\pi), \text{var}(h)=1}{\sup}\hspace{-2em}\{\text{var}[\hspace{-.1em}h(\mathbf{X}^{t+1})\hspace{-.1em}]\hspace{-.2em}-\hspace{-.2em}E[\text{var}[h(\mathbf{X}^{t+1})|\mathbf{X}^t]]\}\right]^{\frac{1}{2}}\notag\\
\hspace{-1.1em}&=&\hspace{-1.2em}\left[1-\hspace{-1em}\underset{h\in L_0^2(\pi), \text{var}(h)=1}{\inf}\{E[\text{var}[h(\mathbf{X}^{t+1})|\mathbf{X}^t]]\}\right]^{\frac{1}{2}}\notag\\
\hspace{-1.1em}&\overset{(b)}{=}&\hspace{-1.2em}\left[1-\hspace{-1em}\underset{h\in L_0^2(\pi), \text{var}(h)=1}{\inf}\hspace{-.3em}\left\{\sum^n_{i=1}\hspace{-.2em}\beta_iE[\text{var}[h(\mathbf{x})|\mathbf{x}_{[-i]}]]\right\}\right]^{\frac{1}{2}}\notag\\
\hspace{-1.1em}&=&\hspace{-1.2em}\left[1-\hspace{-1em}\underset{h\in L_0^2(\pi), \text{var}(h)=1}{\inf}\hspace{-.3em}\left\{\hspace{-.2em}\sum^n_{i=1}\hspace{-.2em}\beta_i\hspace{-.5em}\sum_{\mathbf{x}_{[-i]}}\hspace{-.3em}\text{var}[h(\mathbf{x})|\mathbf{x}_{[-i]}]P(\mathbf{x}_{[-i]})\hspace{-.3em}\right\}\hspace{-.2em}\right]^{\frac{1}{2}}
\label{m5}
\end{eqnarray}
where $(a)$ follows the \emph{law of total variance} of random variable in statistics shown below
{\allowdisplaybreaks\begin{flalign}
\text{var}(\mathbf{A})=E[\text{var}(\mathbf{A}|\mathbf{B})]+\text{var}[E(\mathbf{A}|\mathbf{B})],
\label{xm1}
\end{flalign}}$(b)$ comes from the fact that $\mathbf{X}^t$ and $\mathbf{X}^{t+1}$ differs by only one component $x_i$ and the index $i$ obeys the distribution $\beta_i$'s as a random variable.

On the other hand, because of $\text{var}[x_i|\mathbf{x}_{[-i]}]=\kappa_i>0$ and $\text{var}(h)=1$, it follows that
\begin{equation}
\text{var}[h(\mathbf{x})|\mathbf{x}_{[-i]}]>0,\ \ \forall i
\label{xm1x}
\end{equation}
so as to the existence of the constant lower bound $\kappa^{\dag}$ for the summation
\begin{equation}
\sum^n_{i=1}\text{var}[h(\mathbf{x})|\mathbf{x}_{[-i]}]\geq\kappa^{\dag}>0,
\end{equation}
no matter what $h(\cdot)$ is. By simple induction, the following infimum will be lower bounded as
\begin{equation}
\underset{h\in L_0^2(\pi), \text{var}(h)=1}{\inf}\hspace{-.3em}\left\{\hspace{-.2em}\sum^n_{i=1}\hspace{-.2em}\beta_i\hspace{-.5em}\sum_{\mathbf{x}_{[-i]}}\hspace{-.3em}\text{var}[h(\mathbf{x})|\mathbf{x}_{[-i]}]P(\mathbf{x}_{[-i]})\hspace{-.3em}\right\}=\kappa^{\ddag}>0,
\end{equation}
leading to
\begin{equation}
\text{spec}(\mathbf{F})=(1-\kappa^{\ddag})^{\frac{1}{2}}<1.
\end{equation}

Therefore, by invoking Theorem 2, the proof is completed with $\gamma=1-\text{spec}(\mathbf{F})>0$.
\end{proof}
To summarize, the Markov chain converges exponentially fast to the lattice Gaussian distribution, where the exponential convergence rate $\varrho=\text{spec}(\mathbf{F})$ is derived in (\ref{m5}). Although it is difficult to calculate $\varrho$ explicitly, comprehensive convergence analysis still can be performed, which targets at a smaller $\varrho$. In fact, from Hirschfeld-Gebelein-Rényi (HGR) maximal correlation point of view, $\varrho$ shown in (\ref{xnmb1}) represents the lag-1 maximal correlation between two consecutive Markov states $\mathbf{X}^t$ and $\mathbf{X}^{t+1}$ \cite{LiuFraction}, and Markov states further apart in the chain turns out to be gradually uncorrelated in an exponential way.

\section{Convergence Enhancement for Univariate Sampling}
In this section, Metropolis-within-Gibbs algorithm is proposed for lattice Gaussian sampling. By refining the state space of each univariate sampling, the sampler turns out to be more efficient, resulting in a faster convergence rate.

\subsection{Classical MH Algorithms}
The origin of Metropolis algorithm comes from the celebrated work of \cite{MetropolisOrignial} in 1950's. Then, the original Metropolis algorithm was successfully extended to a more general scheme known as the Metropolis-Hastings (MH) algorithm \cite{Hastings1970}. In particular, let us consider a target invariant distribution $\pi$ together with a proposal distribution
$q(\mathbf{x},\mathbf{y})$. Given the current state $\mathbf{X}^t=\mathbf{x}$ for Markov chain, a state candidate $\mathbf{y}$ for the next Markov move $\mathbf{X}^{t+1}$ is generated from the proposal distribution $q(\mathbf{x},\mathbf{y})$. After that, the acceptance ratio $\alpha$ is computed by
\begin{equation}
\alpha(\mathbf{x},\mathbf{y})=\text{min}\left\{1,\frac{\pi(\mathbf{y})q(\mathbf{y},\mathbf{x})}{\pi(\mathbf{x})q(\mathbf{x},\mathbf{y})}\right\},
\label{quantity compute}
\end{equation}
and $\mathbf{y}$ will be accepted by $\mathbf{X}^{t+1}$ with
probability $\alpha$. Otherwise, $\mathbf{x}$ will be retained by $\mathbf{X}^{t+1}$. In this way, a Markov chain $\{\mathbf{X}^0, \mathbf{X}^1, \ldots\}$ is established with the transition probability $P(\mathbf{X}^t,\mathbf{X}^{t+1})$ as follows:
\begin{equation}
P(\mathbf{X}^t=\mathbf{x},\mathbf{X}^{t+1}=\mathbf{y})=\begin{cases}q(\mathbf{x},\mathbf{y})\alpha(\mathbf{x},\mathbf{y}) \ \ \ \ \ \ \ \ \ \ \ \ \text{if}\ \mathbf{y}\neq\mathbf{x}, \\
       1-\sum_{\mathbf{z}\neq\mathbf{x}}q(\mathbf{x},\mathbf{z})\alpha(\mathbf{x},\mathbf{z})\text{if}\ \mathbf{y}=\mathbf{x}.
       \end{cases}
\label{eqn:RandomDiscrete}
\end{equation}

It is interesting that in MH algorithms, the proposal distribution $q(\mathbf{x},\mathbf{y})$ can be any fixed distribution from which we can conveniently draw samples. To this end, many variations of MH algorithms with different configurations of $q(\mathbf{x},\mathbf{y})$ were proposed.


\subsection{Metropolis-within-Gibbs Algorithm}
In principle, Gibbs algorithm is a special case of MH sampling that tackles with multi-dimensional problems through conditional univariate sampling. More precisely, by letting $q(\mathbf{x},\mathbf{y})=\pi(x_i|\mathbf{x}_{[-i]})$, the acceptance ratio $\alpha$ is always 1 by definition
\begin{equation}
\frac{\pi(\mathbf{y})q(\mathbf{y},\mathbf{x})}{\pi(\mathbf{x})q(\mathbf{x},\mathbf{y})}\hspace{-.2em}=\hspace{-.2em}\frac{\pi(x^{t+1}_i|\mathbf{x}_{[-i]})\pi(\mathbf{x}_{[-i]})\pi(x^{*}_i|\mathbf{x}_{[-i]})}{\pi(x^{*}_i|\mathbf{x}_{[-i]})\pi(\mathbf{x}_{[-i]})\pi(x^{t+1}_i|\mathbf{x}_{[-i]})}\hspace{-.2em}\equiv\hspace{-.2em}1,
\label{v1}
\end{equation}
where $x^{*}_i$ denotes the value of $x^t_i$ for state $\mathbf{x}$. In order to exploit the convergence potential of the univariate sampling in Gibbs algorithm, here we propose the Metropolis-within-Gibbs (MWG) algorithm for lattice Gaussian sampling, which brings uncertainty back to the acceptance-rejection rule in Gibbs algorithm.

\begin{algorithm}[t]
\caption{Metropolis-within-Gibbs Algorithm for Lattice Gaussian Sampling}
\begin{algorithmic}[1]
\Require
$\mathbf{B}, \sigma, \mathbf{c}, \mathbf{X}^0, \beta_i$'s, $t_{\text{mix}}(\epsilon)$
\Ensure
$\mathbf{x}\thicksim \pi,\ \pi\ \text{is within statistical distance of}\ \epsilon\ \text{to}\ D_{\Lambda,\sigma,\mathbf{c}}$
\For {$t=$1,2,\ \ldots}
\State let $\mathbf{x}$ denote the state of $\mathbf{X}^{t-1}$
\State randomly choose index $i$ by distribution $[\beta_1,\ldots,\beta_n]$
\State sample $x_i$ by proposal distribution $q(x_i|\mathbf{x}_{[-i]})$ in (\ref{proposalmwg1})
\State calculate the acceptance quantity $\alpha$ shown in (\ref{proposalmwg})
\State generate a sample $u\sim U[0,1]$
\If {$u\leq \alpha$}
\State get $\mathbf{y}$ with the sampled $x_i$ and let $\mathbf{X}^t=\mathbf{y}$
\Else \ let $\mathbf{X}^t=\mathbf{x}$
\EndIf
\If {$t\geq t_{\text{mix}}(\epsilon)$}
\State output the state of $\mathbf{X}^t$
\EndIf
\EndFor
\end{algorithmic}
\end{algorithm}

In particular, the proposed MWG algorithm can be summarised as the
following three main procedures.

1)\ \ \hspace{-.2em}\emph{Sample from the following univariate proposal distribution to obtain the candidate sample $x^{t+1}_i$},
\begin{equation}
q(\mathbf{x},\mathbf{y}\hspace{-0.1em})\hspace{-0.2em}=\hspace{-0.2em}q(x_i|\mathbf{x}_{[-i]}\hspace{-0.2em})\hspace{-0.2em}=\hspace{-0.2em}\frac{D_{\Lambda,\sigma,\mathbf{c}}(x^{t+1}_i|\mathbf{x}_{[-i]})}{1\hspace{-0.2em}-\hspace{-0.2em}D_{\Lambda,\sigma,\mathbf{c}}(x^{*}_{i}|\mathbf{x}_{[-i]})},
\label{proposalmwg1}
\end{equation}
\emph{where the current} $i$\emph{-th coordinate of }$\mathbf{x}$\emph{, i.e., }$x^{*}_i$\emph{, is eliminated from the sampling candidate space of }$x^{t+1}_i$\emph{ for} $\mathbf{y}$ \emph{to make sure} $x^{t+1}_i\neq x^{*}_i$ and $\mathbf{x}\neq \mathbf{y}$.

2)\ \ \hspace{-.2em}\emph{From} (\ref{quantity compute}), \emph{calculate the acceptance ratio $\alpha(\mathbf{x},\mathbf{y})$}
\begin{equation}
\alpha(\mathbf{x},\mathbf{y})\hspace{-.1em}=\hspace{-.1em}\text{min}\left\{1, \frac{1-D_{\Lambda,\sigma,\mathbf{c}}(x^{*}_i|\mathbf{x}_{[-i]})}{1-D_{\Lambda,\sigma,\mathbf{c}}(x^{t+1}_{i}|\mathbf{x}_{[-i]})}\right\}.
\label{proposalmwg}
\end{equation}

3)\ \ \hspace{-.2em}\emph{Make a decision for $\mathbf{X}^{t+1}$ based on $\alpha(\mathbf{x},\mathbf{y})$ to accept $\mathbf{y}$ or not.}


Different from Gibbs algorithm who always accepts the sampling candidate determinately by $\alpha\equiv1$, a salient feature of MWG algorithm is that the uncertainty arising from the sample acceptance is retained \cite{LiuPeskun} by removing $x^{*}_i$ from the candidate list. To conclude, the proposed MWG algorithm for lattice Gaussian sampling is presented in Algorithm 3.

\subsection{Convergence Rate Analysis}
\begin{my2}
Given the invariant lattice Gaussian distribution $D_{\Lambda,\sigma,\mathbf{c}}$, Metropolis-within-Gibbs algorithm achieves a better exponential convergence performance than Gibbs algorithm by
\end{my2}
\vspace{-1.5em}
\begin{equation}
\varrho_{\text{MWG}}\leq\varrho.
\end{equation}
\begin{proof}
First of all, according to (\ref{eqn:RandomDiscrete}), the transition probability of MWG sampling algorithm is derived as
\begin{equation}
\hspace{-2cm}P_{\text{MWG}}(\mathbf{X}^t=\mathbf{x},\mathbf{X}^{t+1}=\mathbf{y})=q(\mathbf{x},\mathbf{y})\cdot \alpha(\mathbf{x},\mathbf{y})\notag
\end{equation}
\vspace{-1.5em}
\begin{eqnarray}
\hspace{-2.2em}&=&\hspace{-1em}\text{min}\hspace{-.2em}\left\{\hspace{-.3em}\frac{D_{\Lambda,\sigma,\mathbf{c}}(x^{t+1}_i|\mathbf{x}_{[-i]})}{1-D_{\Lambda,\sigma,\mathbf{c}}(x^{*}_{i}|\mathbf{x}_{[-i]})},\frac{D_{\Lambda,\sigma,\mathbf{c}}(x^{t+1}_i|\mathbf{x}_{[-i]})}{1-D_{\Lambda,\sigma,\mathbf{c}}(x^{t+1}_{i}|\mathbf{x}_{[-i]})}\hspace{-.3em}\right\}.
\end{eqnarray}

Compared to the transition probability of Gibbs algorithm given in (\ref{eq:transition}) that
\begin{equation}
P_{i}(x_i|\mathbf{x}_{[-i]})=D_{\Lambda,\sigma,\mathbf{c}}(x^{t+1}_i|\mathbf{x}_{[-i]}),
\end{equation}
it is straightforward to see that
\begin{equation}
P_{\text{MWG}}(\mathbf{X}^t=\mathbf{x},\mathbf{X}^{t+1}=\mathbf{y})\geq P(\mathbf{X}^t=\mathbf{x},\mathbf{X}^{t+1}=\mathbf{y})
\end{equation}
for $\mathbf{x}\neq\mathbf{y}$, which means each off-diagonal element in transition matrix $\mathbf{P}_{\text{MWG}}$ is always larger than that of $\mathbf{P}$. From literatures of MCMC, such a case is known as \emph{Peskun ordering} written by
\begin{equation}
P_{\text{MWG}}(\mathbf{X}^t,\mathbf{X}^{t+1})\succeq P(\mathbf{X}^t,\mathbf{X}^{t+1}).
\label{peskunordering}
\end{equation}

%

Now, we invoke the following Lemma to reveal the relation between Peskun ordering and convergence rate.
\begin{my1}[\hspace{-0.005em}\cite{orderingmcmc}]
Given reversible Markov chains $P$ and $Q$ with stationary distribution $\pi$, if $P\succeq Q$, then their convergence rates satisfy
\end{my1}
\vspace{-1.5em}
\begin{equation}
\varrho_{P}\leq\varrho_{Q}.
\end{equation}

Therefore, according to (\ref{peskunordering}) and Lemma 1, we can immediately obtain that
\begin{equation}
\varrho_{\text{MWG}}\leq\varrho,
\end{equation}
completing the proof.
\end{proof}

The insight behind Peskun ordering is that a Markov chain has smaller probability of remaining in the same position explores the state space more efficiently. Hence, convergence performance is improved by shifting probabilities off the diagonal of the transition matrix, which corresponds to decrease the rejection probability of the proposed moves.

\subsection{Parallel Tempering}
Now, the parallel tempering technique is adopted to the proposed Metropolis-within-Gibbs algorithm to alleviate the possible problems associated with slow mixing Markov chains, which may get stuck during the convergence.

Theoretically, parallel tempering is a generic MCMC sampling method which allows a better convergence. The inspiration of it comes from the idea that a temperature parameter could be used to flatten out the target distribution, thus making the random walk chain for that temperature more likely to mix quickly \cite{ParallelTempering2}. Therefore, according to parallel tempering, the Markov chain induced by the Metropolis-within-Gibbs algorithm for lattice Gaussian sampling can be strengthened as follows.


1)\ \ \hspace{-.2em}\emph{Define a set of target lattice Gaussian distributions $\pi_{t_1},\ldots, \pi_{t_m}$}
\begin{equation}
\pi_{t_j}=D_{\Lambda,t_j\sigma,\mathbf{c}}(\mathbf{x}),\ \ 1\leq j\leq m
\end{equation}
\emph{where} $t_m>\ldots>t_1=1$ \emph{represent different temperature parameters respectively.}


2)\ \ \hspace{-.2em}\emph{Run} $m$ \emph{Markov chains in parallel with the MWG transition probability}
\begin{equation}
\hspace{-2cm}P^j_{\text{MWG}}(\mathbf{X}_j^t=\mathbf{x},\mathbf{X}_j^{t+1}=\mathbf{y})\notag\ \ \ \ \ \ \ \ \ \ \ \ \ \ \ \ \ \ \ \ \
\end{equation}
\vspace{-1.5em}
\begin{eqnarray}
\hspace{-2em}&=&\hspace{-.8em}\text{min}\hspace{-.1em}\left\{\hspace{-.2em}\frac{\pi_{t_j}(x_i|\mathbf{x}_{[-i]})}{1\hspace{-.2em}-\hspace{-.2em}\pi_{t_j}(x^{*}_{i}|\mathbf{x}_{[-i]})},\hspace{-.2em}\frac{\pi_{t_j}(x_i|\mathbf{x}_{[-i]})}{1\hspace{-.2em}-\hspace{-.2em}\pi_{t_j}(x_{i}|\mathbf{x}_{[-i]})}\hspace{-.2em}\right\}
\end{eqnarray}
for $1\leq j\leq m$.

3)\ \ \hspace{-.2em}\emph{After $t_{\text{swap}}$ Markov moves on each Markov chain, consecutively select chain pairs between two neighboring temperatures $t_j$ and $t_{j+1}$, $1\leq j\leq m-1$, then attempt to swap their states with probability}
\begin{equation}
\alpha_{\text{swap}}=\min\left\{1, \frac{\pi_{t_j}(\mathbf{X}^{t+1}_{t_{j+1}})\pi_{t_{j+1}}(\mathbf{X}^{t+1}_{t_j})}{\pi_{t_{j+1}}(\mathbf{X}^{t+1}_{t_{j+1}})\pi_{t_j}(\mathbf{X}^{t+1}_{t_j})}\right\},
\label{tempaaa}
\end{equation}
\emph{otherwise the swap over $\mathbf{X}_j^{t+1}$ and $\mathbf{X}_{j+1}^{t+1}$ is canceled.}

To summarize, this modification essentially allows two types of update. The first one draws samples from distributions $D_{\Lambda,t_j\sigma,\mathbf{c}}(\mathbf{x})$, and the second one is based on a proposal generated from the potential swapping of states between Markov chains. Here, the acceptance probability shown in (\ref{tempaaa}) mainly ensures that the second type of update preserves the stationary distribution \cite{ParallelTempering1}.
Note that only pairs between neighboring temperatures are considered for swapping, where the chances of accepting an exchange are more likely to be higher.

Clearly, with the increase of temperature parameter $t_j$, the lattice Gaussian distribution $D_{\Lambda,t_j\sigma,\mathbf{c}}(\mathbf{x})$ becomes `warm', which would correspond to a uniform distribution over the entire state space. More specifically, the warm distribution mix progressively more rapidly than the cold one which is of primary interest. By allowing the Markov chains to swap states, the convergence performance of the `cold' chain is improved since the state space is traversed more rapidly. Note that such an operation also requires to run multiple chains in parallel, and only the output from one is used as a basis for inference.


\section{Convergence Enhancement for Multivariate Sampling}
To further improve the convergence performance, the blocked strategy, which performs the sampling over multiple components of $\mathbf{x}$ within a block, is investigated. Then, Gibbs-Klein algorithm is proposed for the efficient implementation of blocked Gibbs algorithm.

\subsection{Convergence Analysis of Blocked Sampling}
The idea of blocked strategy in Gibbs algorithm is described by a two-component blocked sampling depicted in Fig. 2. Compared to univariate sampling, by sampling multiple components together, the slow, componentwise moves will be replaced by the fast moves incorporating the information about dependence between components.

\begin{my4}
Given the invariant lattice Gaussian distribution $D_{\Lambda,\sigma,\mathbf{c}}$, the blocked Gibbs algorithm achieves a faster convergence rate than the standard one as
\end{my4}
\vspace{-1.5em}
{\allowdisplaybreaks\begin{flalign}
\varrho_{\text{block}}\leq\varrho.
\label{x3}
\end{flalign}}
\vspace{-1em}
\begin{proof}
First of all, by taking the random index $i$ at each Markov move into account, the term shown in (\ref{xnmb1}) can be described as
{\allowdisplaybreaks\begin{flalign}
\text{var}[E[h(\mathbf{X}^{t+1})|\mathbf{X}^t]]=\sum^n_{i=1}\beta_i\text{var}[E[h(\mathbf{x})|\mathbf{x}_{[-i]}]]
\label{x4}
\end{flalign}}
and subsequently, we have
\vspace{-1em}
\begin{eqnarray}
\varrho\hspace{-.5em}&=&\hspace{-.5em}\hspace{-1.2em}\underset{h\in L_0^2(\pi), \text{var}(h)=1}{\sup}\left[\sum^n_{i=1}\beta_i\text{var}[E[h(\mathbf{x})|\mathbf{x}_{[-i]}]]\right]^{\frac{1}{2}}\notag\\
\hspace{-.5em}&=&\hspace{-.5em}\left[\underset{h\in L_0^2(\pi), \text{var}(h)=1}{\sup}\sum^n_{i=1}\beta_i\text{var}[E[h(\mathbf{x})|\mathbf{x}_{[-i]}]]\right]^{\frac{1}{2}}.
\label{cm2}
\end{eqnarray}

%

For ease of presentation, a two-component blocked sampling scenario is firstly concerned. Typically, suppose components $x_i$ and $x_j$ of $\mathbf{x}$ can be sampled together, then consider the fact that
{\allowdisplaybreaks\begin{flalign}
E[h(\mathbf{x})|\mathbf{x}_{[-i,-j]}]=E[E[h(\mathbf{x})|\mathbf{x}_{[-i]}]|\mathbf{x}_{[-j]}],
\end{flalign}}we can immediately get
{\allowdisplaybreaks\begin{flalign}
\text{var}[E[h(\mathbf{x})|\mathbf{x}_{[-i,-j]}]]\leq\text{var}[E[h(\mathbf{x})|\mathbf{x}_{[-i]}]]
\label{m691}
\end{flalign}}and
{\allowdisplaybreaks\begin{flalign}
\text{var}[E[h(\mathbf{x})|\mathbf{x}_{[-i,-j]}]]\leq\text{var}[E[h(\mathbf{x})|\mathbf{x}_{[-j]}]],
\label{m692}
\end{flalign}}by the law of total variance $\text{var}(\mathbf{A})=E[\text{var}(\mathbf{A}|\mathbf{B})]+\text{var}[E(\mathbf{A}|\mathbf{B})]$.

Therefore, given the index selection probabilities $\beta_i$ and $\beta_j$, we have
\begin{eqnarray}
\hspace{-2em}(\beta_i\hspace{-.3em}+\hspace{-.3em}\beta_j)\text{var}[E[h(\mathbf{x})|\mathbf{x}_{[-i,-j]}]]\hspace{-.8em}&\leq&\hspace{-.8em}\beta_i\text{var}[E[h(\mathbf{x})|\mathbf{x}_{[-i]}]]\notag\\
\hspace{-1em}&&\hspace{-.5em}+\beta_j\text{var}[E[h(\mathbf{x})|\mathbf{x}_{[-j]}]].
\label{m69}
\end{eqnarray}
From (\ref{cm2}), this indicates a more efficient convergence rate $\rho$ for the blocked sampling over $x_i$ and $x_j$.

Inductively, this two-component blocked sampling over coordinates $i$ and $j$ can be easily extended to any larger size blocked sampling. Hence, according to (\ref{cm2}) and (\ref{m69}), it follows that
\begin{equation}
\varrho_{\text{block}}\leq\varrho,
\end{equation}
completing the proof.
\end{proof}

From (\ref{m691}), it is straightforward to check that the convergence performance also improves gradually by grouping more elements into the block
{\allowdisplaybreaks\begin{flalign}
\text{var}[E[h(\mathbf{x})|\mathbf{x}_{[-\text{block},-j]}]]\leq\text{var}[E[h(\mathbf{x})|\mathbf{x}_{[-\text{block}]}]]
\end{flalign}}since a larger block size allows
moves in more general directions. If all the components forming
a single block could be sampled directly, there would be
no need for MCMC sampling. In this regard, blocked
strategy is strongly recommended if sampling over multivariate can be efficiently performed.

\newcounter{TempEqCnt}                         
\setcounter{TempEqCnt}{\value{equation}} 
\setcounter{equation}{56}
\begin{figure*}[bb]
\hrulefill
\begin{eqnarray}
P(\mathbf{z}_{\text{block}}\mid\mathbf{z}_{[-\text{block}]})&=&\prod^m_{i=1}D_{\mathbb{Z},\sigma_{m+1-i},\widetilde{z}_{m+1-i}}(z_{m+1-i})\notag\\
&=&\frac{e^{-\frac{1}{2\sigma^2}\sum_{i=1}^m\left(\overline{c}_{m+1-i}-\sum_{j=m+1-i}^mr_{m+1-i,j}z_j\right)^2}}{\prod^m_{i=1}\sum_{z_{m+1-i}\in \mathbb{Z}}e^{-\frac{1}{2\sigma^2}\left(\overline{c}_{m+1-i}-\sum_{j=m+1-i}^mr_{m+1-i,j}z_j\right)^2}}\notag\\
&=&\frac{e^{-\frac{1}{2\sigma^2}\parallel\overline{\mathbf{c}}-\overline{\mathbf{r}}\mathbf{z}_{\text{block}}\parallel^2}}{\prod^m_{i=1}\sum_{z_{m+1-i}\in \mathbb{Z}}e^{-\frac{1}{2\sigma^2}\left(r_{m+1-i,m+1-i}z_{m+1-i}-\overline{c}_{m+1-i}+\sum_{j=m+2-i}^mr_{m+1-i,j}z_j\right)^2}}\notag\\
&=&\frac{\rho_{\mathcal{L}(\overline{\mathbf{r}}),\sigma,\overline{\mathbf{c}}}(\mathbf{z}_{\text{block}})}{\prod^m_{i=1}\rho_{\sigma}(r_{m+1-i,m+1-i}\mathbb{Z}+\xi)},
\label{1 conditional distribution a}
\end{eqnarray}
\vspace*{4pt}
\end{figure*}
\setcounter{equation}{\value{TempEqCnt}}

\subsection{Efficient Blocked Sampling by Gibbs-Klein Algorithm}
Although blocked sampling achieves a better convergence rate than univariate one, sampling over a block is generally more costly than componentwise sampling because its sampling space increases exponentially with the block size. Because of this, we propose to use Klein's algorithm for multi-component sampling, which leads to the Gibbs-Klein algorithm.

At each step of Markov chain, the proposed Gibbs-Klein algorithm randomly picks up $m$ components of $\mathbf{x}$ to update. For a better illustration of the proposed sampling, here we establish another new scheme but equivalent to the foregoing one, which resorts to the help of permutation matrices. In particular, an $n \times n$ permutation matrix $\mathbf{E}$ is applied to sort the updating order within the blocked sampling. If $\mathbf{E}$ is randomly generated, then Gibbs-Klein algorithm on $m$ randomly chosen components will be equivalent to sample $m$ consecutive components of $\mathbf{z}$ in a fixed order, where $\mathbf{z}=\mathbf{E^{-1}x}$ and $\mathbf{\widetilde{B}}=\mathbf{BE}$. Here, the selection probabilities $[\beta_1, \ldots, \beta_n]$ are set as uniform to make it simple.
For simplicity, we always consider the block formed by the first $m$ components of $\mathbf{z}$, namely $\mathbf{z}_{\text{block}}=[z_1,\ldots,z_m]^T$. Then, it follows that,
\begin{eqnarray}
\hspace{-1em}D_{\mathcal{L}(\mathbf{B}),\sigma,\mathbf{c}}(\mathbf{x})\hspace{-.8em}&=&\hspace{-.8em}\frac{e^{-\frac{1}{2\sigma^2}\parallel \mathbf{Bx}-\mathbf{c} \parallel^2}}{\sum_{\mathbf{x} \in \mathbb{Z}^n}e^{-\frac{1}{2\sigma^2}\parallel \mathbf{Bx}-\mathbf{c} \parallel^2}}\notag\\
\hspace{-.8em}&=&\hspace{-.8em}\frac{e^{-\frac{1}{2\sigma^2}\parallel \widetilde{\mathbf{B}}\mathbf{z}-\mathbf{c} \parallel^2}}{\sum_{\mathbf{z} \in \mathbb{Z}^n}e^{-\frac{1}{2\sigma^2}\parallel \widetilde{\mathbf{B}}\mathbf{z}-\mathbf{c} \parallel^2}}=D_{\mathcal{L}(\widetilde{\mathbf{B}}),\sigma,\mathbf{c}}(\mathbf{z}).
\end{eqnarray}

After QR-decomposition
$\mathbf{\widetilde{B}}=\mathbf{QR}$ and calculating $\mathbf{c}'=\mathbf{Q}^{T}\mathbf{c}$, $z_i$ in the block is sampled from the following 1-dimensional distribution with the backward order from $z_m$ to $z_1$:
\begin{eqnarray}
P_i(z_i|\mathbf{\overline{z}}_{[-i]})&=&D_{\mathbb{Z},\sigma_i,\widetilde{z}_i},
\label{1 conditional distribution zzz}
\end{eqnarray}
where $\sigma_i=\frac{\sigma}{| r_{i,i}|}$, $ \mathbf{\overline{z}}_{[-i]} =[z_{i+1},\ldots,z_m,z_{m+1},\ldots,z_{n}]^T$ and $\widetilde{z}_i=\frac{c'_i-\sum^m_{j=i+1}r_{i,j}z_j-\sum^n_{j{'}=m+1}r_{i,j{'}}z_{j{'}}}{r_{i,i}}$. Algorithm 4 gives the proposed Gibbs-Klein algorithm, where $\mathbf{z}=[\mathbf{z}_{\text{block}};\mathbf{z}_{[-\text{block}]}]$ and $\mathbf{z}_{[-\text{block}]}=[z_{m+1},\ldots,z_{n}]^T$. Then, the Markov state of $\mathbf{X}^t$ is obtained by the transformation $\mathbf{X}^t=\mathbf{x}=\mathbf{Ez}$.

Based on Algorithm 4, the extension to other scan strategies is possible. We point out that the implementation via the random permutation matrix $\mathbf{E}$ is not that efficient due to the usage of QR decompositions, but could offer a straightforward way to illustrate and analyze Gibbs-Klein algorithm. In other words, Gibbs-Klein algorithm can be carried out without QR decomposition.

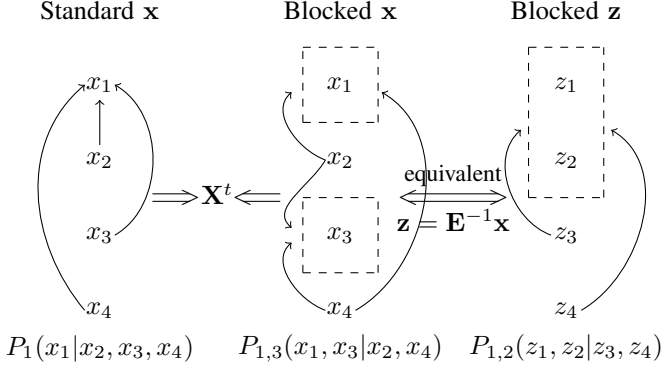
\begin{figure}[t]
\begin{center}
\begin{tikzpicture}
\node  at (0,4) {$x_1$};
\node  at (0,3) {$x_2$};
\node  at (0,2) {$x_3$};
\node  at (0,1) {$x_4$};
\node  at (0,0.5) {$P_1(x_1|x_2,x_3,x_4)$};
\node  at (0,5) {Standard $\mathbf{x}$};
\node  at (3.2,5) {Blocked $\mathbf{x}$};
\node  at (6.2,5) {Blocked $\mathbf{z}$};
\draw [->] (0.2,2) to [out=30,in=330] (0.2,4);
\draw [->] (-0.2,1) to [out=130,in=220] (-0.2,4);
\draw [->](0,3.2) --(0,3.8);
\draw (0.7,2.55) --(1.2,2.55);
\draw (0.7,2.45) --(1.2,2.45);
\node  at (1.2,2.5) {$>$};
\node  at (1.5,2.55) {\ $\mathbf{X}^{t}$};
\draw (1.9,2.55) --(2.4,2.55);
\draw (1.9,2.45) --(2.4,2.45);
\node  at (1.9,2.5) {$<$};
\node  at (3.2,4) {$x_1$};
\node  at (3.2,3) {$x_2$};
\node  at (3.2,2) {$x_3$};
\node  at (3.2,1) {$x_4$};
\node  at (3.2,0.5) {$P_{1,3}(x_1,x_3|x_2,x_4)$};
\draw [dashed] (2.7, 1.5) -- (3.7, 1.5);
\draw [dashed] (2.7, 2.5) -- (3.7, 2.5);
\draw [dashed] (2.7, 3.5) -- (3.7, 3.5);
\draw [dashed] (2.7, 4.5) -- (3.7, 4.5);
\draw [dashed] (2.7, 1.5) -- (2.7, 2.5);
\draw [dashed] (2.7, 3.5) -- (2.7, 4.5);
\draw [dashed] (3.7, 1.5) -- (3.7, 2.5);
\draw [dashed] (3.7, 3.5) -- (3.7, 4.5);
\draw [->] (3.4,1) to [out=30,in=330] (3.75,3.9);
\draw [->] (3,1) to [out=160,in=220] (2.55,1.9);
\draw [->] (3,3) to [out=160,in=220] (2.55,3.9);
\draw [->] (3,3) to [out=240,in=140] (2.55,2.1);
\draw (4.1,2.55) --(5.3,2.55);
\draw (4.1,2.45) --(5.3,2.45);
\node  at (5.3,2.5) {$>$};
\node  at (4.1,2.5) {$<$};
\node  at (4.7,2.8) {\small{equivalent}};
\node  at (4.7,2.2) {$\mathbf{z}=\mathbf{E}^{-1}\mathbf{x}$};
\node  at (6.2,4) {$z_1$};
\node  at (6.2,3) {$z_2$};
\node  at (6.2,2) {$z_3$};
\node  at (6.2,1) {$z_4$};
\node  at (6.2,0.5) {$P_{1,2}(z_1,z_2|z_3,z_4)$};
\draw [dashed] (5.7, 2.5) -- (6.7, 2.5);
\draw [dashed] (5.7, 4.5) -- (6.7, 4.5);
\draw [dashed] (5.7, 2.5) -- (5.7, 4.5);
\draw [dashed] (6.7, 2.5) -- (6.7, 4.5);
\draw [->] (6.4,1) to [out=30,in=330] (6.8,3.4);
\draw [->] (6,2) to [out=160,in=220] (5.6,3.4);
\end{tikzpicture}
\end{center}
\caption{Illustration of standard and blocked Gibbs sampling strategies. Components within the dashed block are sampled as a whole by blocked Gibbs algorithm, where the usage of permutation matrix $\mathbf{E}$ facilitates Klein's algorithm by forming the block with the first $2$ components of $\mathbf{z}$.}
  \label{Graphical representations}
\end{figure}

\subsection{Validity of Gibbs-Klein Algorithm}
Now, the validity of Gibbs-Klein algorithm is verified by showing its ergodicity, where rejection sampling is resorted to make sure the generated distribution by Gibbs-Klein is exact lattice Gaussian distribution.


\begin{my1}
For a given invariant lattice Gaussian distribution $D_{\Lambda,\sigma,\mathbf{c}}$, if $\sigma \geq\omega(\sqrt{\text{log}\ m})\cdot\max_{1\leq i\leq m}|r_{i,i}|$, the blocked sampling probability $P(\mathbf{z}_{\text{block}}\mid\mathbf{z}_{[-\text{block}]})$ in Gibbs-Klein algorithm follows the distribution
\end{my1}
\vspace{-1em}
\begin{equation}
D_{\hspace{-.2em}\mathcal{L}(\widetilde{\mathbf{B}}),\sigma,\mathbf{c}}(\hspace{-.1em}\mathbf{z}_{\text{block}}|\mathbf{z}_{[-\text{block}]}\hspace{-.1em})\hspace{-.2em}=\hspace{-.2em}\frac{e^{-\frac{1}{2\sigma^2}\parallel \mathbf{\widetilde{B}z}-\mathbf{c} \parallel^2}}{\sum_{\mathbf{z}_{\text{block}} \in \mathbb{Z}^m}\hspace{-.2em}e^{-\frac{1}{2\sigma^2}\parallel \mathbf{\widetilde{B}z}-\mathbf{c} \parallel^2}},
\label{sampling radius bound}
\end{equation}
\begin{flushleft}
\emph{where} $\mathbf{z}=[\mathbf{z}_{\text{block}};\mathbf{z}_{[-\text{block}]}]$.
\end{flushleft}
\begin{proof}
From (\ref{1 conditional distribution zzz}) and by induction, the blocked sampling probability $\mathbf{z}_{\text{block}}$ conditioned on $\mathbf{z}_{[-\text{block}]}$ is given by
\begin{equation}
P(\mathbf{z}_{\text{block}}\mid\mathbf{z}_{[-\text{block}]})=\prod^m_{i=1}P(z_{m+1-i}|\mathbf{\overline{z}}_{[-(m+1-i)]}).
\label{1 conditional distribution zzzz}
\end{equation}

Then, according to (\ref{1 conditional distribution zzz}) and (\ref{1 conditional distribution zzzz}), we have the derivation shown in (\ref{1 conditional distribution a}), where $\overline{c}_i=c'_i-\sum_{j^{'}=m+1}^nr_{i,j^{'}}z_{j^{'}}$, $\overline{\mathbf{c}}=[\overline{c}_1,\ldots, \overline{c}_m]^T$, $\xi=\sum_{j=m+2-i}^mr_{m+1-i,j}z_j-\overline{c}_{m-i+i}$ and $\overline{\mathbf{r}}$ is the $m \times m$ segment of $\mathbf{R}$ with $r_{1,1}$ to $r_{m,m}$ in diagonal. Clearly, the effect of the subvector $\mathbf{z}_{[-\text{block}]}$ is hidden in $\overline{c}_i$.

In \cite{Regevlearning}, it has been demonstrated  that if $\sigma>\eta_\varepsilon(\mathcal{L}(\overline{\mathbf{r}}))$, then
\setcounter{equation}{57}
\begin{equation}
\frac{\prod^m_{i=1}\rho_{\sigma}(r_{i,i}\mathbb{Z}+\xi)}{\prod^m_{i=1}\rho_{\sigma}(r_{i,i}\mathbb{Z})}\in\left(\left(\frac{1-\varepsilon}{1+\varepsilon}\right)^m,1\right]
\label{lemma 1}
\end{equation}
which means $\prod^m_{i=1}\rho_{\sigma}(r_{i,i}\mathbb{Z}+\xi)$ can be substituted by $\prod^m_{i=1}\rho_{\sigma}(r_{i,i}\mathbb{Z})$ within negligible errors when $\varepsilon$ is sufficiently small.

As shown in \cite{Trapdoor}, $\eta_\varepsilon(\Lambda)$ with negligible $\varepsilon$ is upper bounded as $\eta_\varepsilon(\Lambda)\leq \omega(\sqrt{\text{log}\ n})\cdot\text{max}_{1\leq i\leq n}\|\mathbf{\widehat{b}}_i\|$. Therefore, if $\sigma\geq \omega(\sqrt{\text{log}\ m})\cdot\text{max}_{1\leq i\leq m}\|r_{i,i}\|$, $P(\mathbf{z}_{\text{block}}\mid\mathbf{z}_{[-\text{block}]})$ shown in (\ref{1 conditional distribution a}) can be rewritten as
\begin{equation}
P(\mathbf{z}_{\text{block}}\hspace{-0.3em}\mid\hspace{-0.3em}\mathbf{z}_{[-\text{block}]})\simeq\frac{\rho_{\mathcal{L}(\overline{\mathbf{r}}),\sigma,\overline{\mathbf{c}}}(\mathbf{z}_{\text{block}})}{\prod^m_{i=1}\rho_{\sigma}(r_{i,i}\mathbb{Z})},
\end{equation}
where ``$\simeq$" represents equality up to a negligible error. Because the denominator is independent of $\mathbf{z}_{\text{block}}$, $\mathbf{z}_{[-\text{block}]}$ and $\mathbf{c}$, it can be viewed as a constant.

Here, in order to remove the latent negligible errors shown above, the classic rejection sampling can be applied to yield an exact sample (see \cite{BrakerskiLPRS13} for more details). Specifically, the candidate of $\mathbf{z}_{\text{block}}$ is outputted with probability
\begin{equation}
\alpha_{\text{accept}}=\frac{\prod^m_{i=1}\rho_{\sigma}(r_{i,i}\mathbb{Z}+\xi)}{\prod^m_{i=1}\rho_{\sigma}(r_{i,i}\mathbb{Z})}
\label{mp1}
\end{equation}
and this probability can be efficiently computed (achieve any desired $t$ bits of accuracy in time $poly(t)$, $t$ denotes the number of iterations), as shown in \cite{BrakerskiLPRS13}.
Therefore, under the help of rejection sampling, it follows that
\begin{equation}
P(\mathbf{z}_{\text{block}}\hspace{-0.3em}\mid\hspace{-0.3em}\mathbf{z}_{[-\text{block}]})=\frac{\rho_{\mathcal{L}(\overline{\mathbf{r}}),\sigma,\overline{\mathbf{c}}}(\mathbf{z}_{\text{block}})}{\prod^m_{i=1}\rho_{\sigma}(r_{i,i}\mathbb{Z})},
\end{equation}
and the output has a lattice Gaussian distribution $D_{\mathcal{L}(\overline{\mathbf{r}}),\sigma,\overline{\mathbf{c}}}(\mathbf{z}_{\text{block}})=D_{\mathcal{L}(\widetilde{\mathbf{B}}),\sigma,\mathbf{c}}(\mathbf{z}_{\text{block}}|\mathbf{z}_{[-\text{block}]})$.
\end{proof}

Because $D_{\mathcal{L}(\widetilde{\mathbf{B}}),\sigma,\mathbf{c}}(\mathbf{z})$ and $D_{\Lambda,\sigma,\mathbf{c}}(\mathbf{x})$ describe the same lattice Gaussian distribution, namely,
\begin{equation}
\frac{e^{-\frac{1}{2\sigma^2}\parallel \mathbf{\widetilde{B}z}-\mathbf{c} \parallel^2}}{\sum_{\mathbf{z}_{\text{block}} \in \mathbb{Z}^m}\hspace{-.2em}e^{-\frac{1}{2\sigma^2}\parallel \mathbf{\widetilde{B}z}-\mathbf{c} \parallel^2}}\hspace{-.2em}\triangleq\hspace{-.2em}\frac{e^{-\frac{1}{2\sigma^2}\parallel \mathbf{B}\mathbf{x}-\mathbf{c} \parallel^2}}{\sum_{\mathbf{x}_{\text{block}} \in \mathbb{Z}^m}\hspace{-.2em}e^{-\frac{1}{2\sigma^2}\parallel \mathbf{Bx}-\mathbf{c} \parallel^2}}
\label{sampling radius bound}
\end{equation}
$D_{\mathcal{L}(\widetilde{\mathbf{B}}),\sigma,\mathbf{c}}(\mathbf{z}_{\text{block}}|\mathbf{z}_{[-\text{block}]})$ actually is one-to-one correspondence with $D_{\mathcal{L}(\mathbf{B}),\sigma,\mathbf{c}}(\mathbf{x}_{\text{block}}|\mathbf{x}_{[-\text{block}]})$. Therefore, according to Lemma 2, Gibbs-Klein algorithm is capable to sample multiple components of $\mathbf{x}$ at each Markov move. We then arrive at the following Theorem.

\begin{algorithm}[t]
\caption{Gibbs-Klein Algorithm for Lattice Gaussian Sampling}
\begin{algorithmic}[1]
\Require
$\mathbf{B}, \sigma, \mathbf{c}, \mathbf{X}^0, t_{\text{mix}}(\epsilon)$;
\Ensure
$\mathbf{x}\thicksim \pi,\ \pi\ \text{is within statistical distance of}\ \epsilon\ \text{to}\ D_{\Lambda,\sigma,\mathbf{c}}$
\For {$t=$1,2\ \ldots}
\State let $\mathbf{x}$ denote the state of $\mathbf{X}^{t-1}$
\State randomly generate a permutation matrix $\mathbf{E}$
\State let $\mathbf{\widetilde{B}}=\mathbf{BE}$ and $\mathbf{z}=\mathbf{E}^{-1}\mathbf{x}$
\State let $\mathbf{\widetilde{B}}=\mathbf{QR}$ and $\mathbf{c}'=\mathbf{Q}^{T}\mathbf{c}$
\For {$k=1$,\ \ldots}
\For {$i=m$,\ \ldots,\ 1}
\State let $\sigma_i=\frac{\sigma}{|r_{i,i}|}$
\State let $\widetilde{z}_i\hspace{-0.3em}=\hspace{-0.3em}\frac{c'_i-\sum^m_{j=i+1}r_{i,j}z_j-\sum^n_{j^{'}=m+1}r_{i,j^{'}}z_{j^{'}}}{r_{i,i}}$
\State sample $z_i$ from $D_{\mathbb{Z},\beta_i,\widetilde{z}_i}$
\EndFor
\State calculate the acceptance ratio $\alpha_{\text{accept}}$ shown in (\ref{mp1})
\State generate a sample $u\sim U[0,1]$
\If {$u\leq \alpha_{\text{accept}}$}
\State output $\mathbf{z}_{\text{block}}$ as the exact sample from $D_{\mathcal{L}(\overline{\mathbf{r}}),\sigma,\overline{\mathbf{c}}}$
\State \textbf{Break}
\EndIf
\EndFor
\State update $\mathbf{z}=[\mathbf{z}_{\text{block}};\mathbf{z}_{[-\text{block}]}]^T$
\State return $\mathbf{x}=\mathbf{Ez}$ and let $\mathbf{X}^t=\mathbf{x}$
\If {$t\geq t_{\text{mix}}(\epsilon)$}
\State output the state of $\mathbf{X}^t$
\EndIf
\EndFor
\end{algorithmic}
\end{algorithm}

\begin{my2}
For $\sigma \geq \omega(\sqrt{\log  m})\cdot\max_{1\leq i\leq m}|r_{i,i}|$, the Markov chain induced by the Gibbs-Klein algorithm with block size $m$ is ergodicity with respect to the lattice Gaussian distribution $D_{\Lambda,\sigma,\mathbf{c}}$ as
\begin{equation}
\underset{t\rightarrow\infty}{\lim}\|P^t(\mathbf{x}, \cdot)-D_{\Lambda,\sigma,\mathbf{c}}\|_{TV}=0
\end{equation}
for all states $\mathbf{x}\in\mathbb{Z}^n$.
\end{my2}

\begin{proof}
Based on Definition 1, we now prove the ergodicity by verifying the reversibility, irreducibility and aperiodic of the underlying Markov chain.

To start with, let $\mathbf{x}_i$ and $\mathbf{x}_j$ be two adjacent states in Gibbs-Klein algorithm. For block size $m$, every two adjacent states in Gibbs-Klein algorithm differ from each other by at most $m$ components.
For convenience, we express them as
\begin{equation}
\mathbf{x}_i=[\mathbf{x}_{\text{block}(i)},\mathbf{x}_{[-\text{block}]}]
\end{equation}
and
\begin{equation}
\mathbf{x}_j=[\mathbf{x}_{\text{block}(j)},\mathbf{x}_{[-\text{block}]}],
\end{equation}
where $\mathbf{x}_{\text{block}(i)}$ and $\mathbf{x}_{\text{block}(j)}$ denote the $m$ components belonging to $\mathbf{x}_i$ and $\mathbf{x}_j$, respectively. Then, the transition probability of Gibbs-Klein algorithm is
{\allowdisplaybreaks\begin{flalign}
P(\mathbf{X}^t\hspace{-.3em}=\hspace{-.2em}\mathbf{x}_i, \hspace{-.2em} \mathbf{X}^{t+1}\hspace{-.3em}=\hspace{-.2em}\mathbf{x}_j)=&P(\mathbf{x}_{\text{block}(i)}\rightarrow\mathbf{x}_{\text{block}(j)}|\mathbf{x}_{[-\text{block}]}) \notag\\
\overset{(c)}{=}&P(\mathbf{x}_{\text{block}(j)}|\mathbf{x}_{[-\text{block}]})\notag\\
=&\frac{e^{-\frac{1}{2\sigma^2}\parallel \mathbf{B}\mathbf{x}_j-\mathbf{c} \parallel^2}}{\sum_{\mathbf{x}_{\text{block}} \in \mathbb{Z}^m}e^{-\frac{1}{2\sigma^2}\parallel \mathbf{Bx}-\mathbf{c} \parallel^2}},
\label{bbbb}
\end{flalign}}where $(c)$ is due to the fact that $\mathbf{x}_{\text{block}}$ is sampled only conditioned on $\mathbf{x}_{[-\text{block}]}$.


To show the Markov chain is irreducible, note that given a state $\mathbf{x}_i$ one can attain with positive probability in one step
any state $\mathbf{x}_j$ which shares $>= (n-m)$ components with $\mathbf{x}_i$. Now, if $\mathbf{x}_i$ and $\mathbf{x}_j$ have, say, $d < n-m$ components in common, there is
always a positive probability that after each step they get exactly one more component in common. So we can go in $n-d$ steps from one to the other. But as soon as $m >= 2$, we can assume that at the first step we get two more components in common, and then one at each further step, so we can go with positive probability in $n-d-1$ steps.

On the other hand, it is clear to see that the number of steps required to move between any two states (can be the same state) is arbitrary without any limitation to be a multiple of some integer. Put another way, the chain is not forced into some cycle with fixed period between certain states. Therefore, the Markov chain is aperiodic.

\begin{figure}[t]
\includegraphics[width=3.5in]{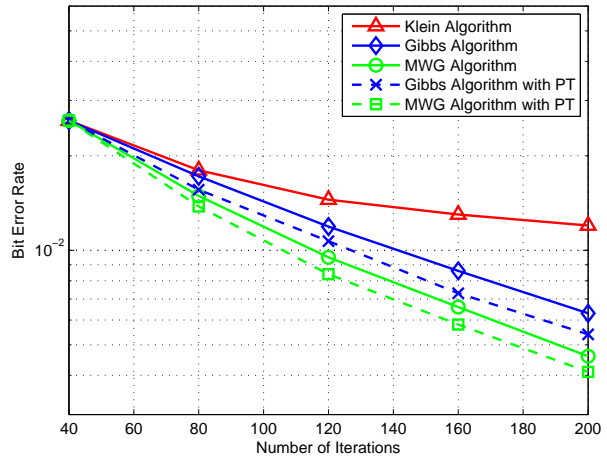}
\vspace{-1em}
  \caption{Bit error rate versus the number of iterations for the uncoded $6 \times 6$ MIMO system using 16-QAM.}
  \label{simulation 1}
\end{figure}

As for reversibility, it is no hard to check that the following relationship holds
\begin{equation}
D_{\Lambda,\sigma,\mathbf{c}}(\mathbf{x}_i)P(\mathbf{x}_i, \mathbf{x}_j)= D_{\Lambda,\sigma,\mathbf{c}}(\mathbf{x}_j)P(\mathbf{x}_j, \mathbf{x}_i)
\end{equation}
with the same expression
\begin{equation}
\frac{e^{-\frac{1}{2\sigma^2}\parallel \mathbf{B}\mathbf{x}_i-\mathbf{c} \parallel^2}}{\sum_{\mathbf{x} \in \mathbb{Z}^n}e^{-\frac{1}{2\sigma^2}\parallel \mathbf{Bx}-\mathbf{c} \parallel^2}}\cdot\frac{e^{-\frac{1}{2\sigma^2}\parallel \mathbf{B}\mathbf{x}_j-\mathbf{c} \parallel^2}}{\sum_{\mathbf{x}_{\text{block}} \in \mathbb{Z}^m}e^{-\frac{1}{2\sigma^2}\parallel \mathbf{Bx}-\mathbf{c} \parallel^2}}.
\end{equation}
Thus, the conclusion follows, completing the proof.
\end{proof}

After showing the Markov chain induced by Gibbs-Klein algorithm is ergodic, it is straightforward to arrive at the geometric ergodicity by the same argument of Corollary 1, and its proof is omitted here.
\begin{my7}
Given the invariant lattice Gaussian distribution $D_{\Lambda,\sigma,\mathbf{c}}$, the Markov chain induced by Gibbs-Klein algorithm is geometrically ergodic
\end{my7}
\vspace{-1em}
\begin{equation}
\|P^t(\mathbf{x}, \cdot)-D_{\Lambda,\sigma,\mathbf{c}}\|_{TV}\leq M(\mathbf{x})\varrho^t_{\text{block}}
\label{geo-ergodic}
\end{equation}
\emph{with} $\varrho_{\text{block}}\leq\varrho$.

Note that the block size is determined by the given standard deviation $\sigma$, the larger $\sigma$, the larger block size $m$ and faster convergence rate $\varrho_{\text{block}}$. This is in agreement with the fact we knew before: if $\sigma$ is sufficiently large, then there is no need of MCMC for lattice Gaussian sampling since Klein's algorithm can be applied directly with polynomial complexity.

\begin{figure}[t]
\includegraphics[width=3.5in]{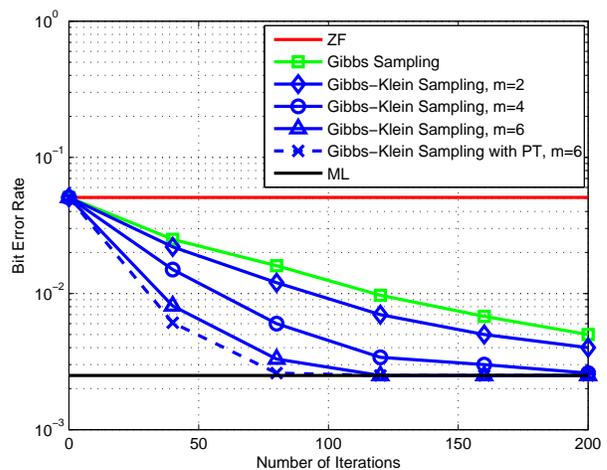}
\vspace{-1em}
  \caption{Bit error rate versus the number of iterations for the uncoded $4 \times 4$ MIMO system using 16-QAM.}
  \label{simulation 1}
\end{figure}

\section{Simulation Results}
In this section, the performance of Gibbs-based sampling schemes for lattice Gaussian distribution are exemplified in the context of MIMO detection.

Specifically, simulation results for an $n\times n$ MIMO system with a square channel matrix containing i.i.d. Gaussian entries are presented. The $i$-th entry of the transmitted signal $\mathbf{x}$, denoted as $x_i$, is a modulation symbol taken independently from a $Q^2$-QAM constellation $\mathcal{X}\in\mathbb{Z}$ with Gray mapping. Meanwhile, it is assumed a flat fading environment, where the channel matrix $\mathbf{H}$ contains uncorrelated complex Gaussian fading gains with unit variance and remains constant over each frame duration. Let $E_b$ represents the average power per bit at the receiver, then $E_b/N_0=n/(\text{log}_2(M)\sigma_w^2)$ holds where $M$ is the modulation level and $\sigma_w^2$ is the noise power\footnote{In \cite{HassibiMCMCnew}, the noise variance $\sigma_w^2$ is used as the sampling variance over the discrete Gaussian distribution, but this would lead to a stalling problem at high SNRs \cite{MCMCKumarGibbsStalling}.}. Then, we can construct the system model as
\begin{equation}
\mathbf{c}=\mathbf{H}\mathbf{x}+\mathbf{w},
\label{eqn:System Model3}
\end{equation}
and this decoding problem of $\widehat{\mathbf{x}}=\underset{\mathbf{x}\in \mathcal{X}^{n}}{\operatorname{arg~min}} \, \|\mathbf{c}-\mathbf{H}\mathbf{x}\|^2$ can be solved by sampling over the discrete Gaussian distribution
\begin{equation}
P_{\mathcal{L}(\mathbf{H}),\sigma,\mathbf{c}}(\mathbf{x})=\frac{e^{-\frac{1}{2\sigma^2}\parallel \mathbf{Hx}-\mathbf{c} \parallel^2}}{\sum_{\mathbf{x} \in \mathcal{X}^n}e^{-\frac{1}{2\sigma^2}\parallel \mathbf{Hx}-\mathbf{c} \parallel^2}}
\label{discrete gaussian distribution}
\end{equation}
because the optimal solution has the largest probability making it most likely be encountered by sampling (this complex decoding system is straightforward to be extended to the real-valued system \cite{CongRandom,xia1}). For this reason, we examine the decoding error probabilities to approximately compare the convergence rates of Markov chains.
Meanwhile, given the definition of geometric ergodicity shown in (\ref{geo-ergodic}), the choice of the starting state $\mathbf{x}$ also has an impact upon the convergence performance. In theory, it could be any candidate from the state space but a poor choice may intensively increase the required mixing time. To this end, starting the Markov chain with $\mathbf{x}$ as close to the center of the distribution as possible would be a judicious choice. As a simple solution, Babai rounding algorithm (also known as zero-forcing decoding) is recommended here to output the suboptimal result for the initial Markov state \cite{Babai}.

Fig. 3 depicts the bit error rates (BER) of the different sampling schemes in a $6\times6$ uncoded MIMO system with 16-QAM. This corresponds to lattice dimension $n=12$. For a fair comparison, we follow Klein's choice of $\sigma=\text{min}_{1\leq i\leq n}\|\mathbf{\widehat{b}}_i\|/\sqrt{\text{log}\ n}$ and run the univariate sampling in both MWG and Gibbs algorithm for $n$ times as a full iteration. Additionally, the parallel tempering technique that fastens the mixing by utilizing the tuning temperatures is also illustrated. For the consideration of computational complexity, only two markov chains are applied for parallel tempering with $t_1=1$ and $t_2=2$, where the swap gap $t_{\text{swap}}$ is set as 1. As shown in Fig. 3, the decoding performance improves with the number of iterations. In particular, Klein's sampling is not as good as MCMC sampling schemes since it does not really produce Gaussian samples \cite{Trapdoor}. On the other hand, as demonstrated, the proposed Metropolis-within-Gibbs algorithm outperforms Gibbs algorithm under the same number of iterations, implying a better convergence performance. Meanwhile, parallel tempering  is strongly recommended if parallel implementation is allowed.

In Fig. 4 illustrates the bit error rates (BER) decoding performance by Gibbs-based multivariate sampling over lattice Gaussian distribution, and its enhancement result by parallel tempering is also given. Specifically, in a $4\times4$ uncoded MIMO system with 16-QAM, for a fair comparison, when the block size is $m$, we run block sampling for $n/m$ times, and count this as a full iteration for Gibbs-Klein algorithm. This corresponds to lattice dimension $n=8$. As can be seen clearly, with the same number of iterations, the decoding performance improves with the block size, which indicates a faster convergence rate. These multiple updates are still dictated by the conditional lattice Gaussian distribution, which takes the correlation structure into account. In this regard, block technique is worth trying for sampling decoding to enhance its performance.

\section{Conclusion}
In this paper, the classic Gibbs algorithm for lattice Gaussian sampling is studied in full details. By resorting to spectral radius of the forward operator, a comprehensive analysis is conducted to prove the geometric ergocidity of Gibbs algorithm for lattice Gaussian sampling, which means the underlying Markov chain converges to the lattice Gaussian distribution in an exponential way. Moreover, by showing the spectral radius of the forward operator exactly characterizes the convergence rate, analysis and optimization are performed to further enhance the convergence performance, where Metropolis-within-Gibbs (MWG) and Gibbs-Klein (GK) algorithms for univariate and multivariate sampling are proposed respectively. Moveover, the validity of Gibbs-Klein algorithm for blocked sampling is confirmed by ergodicity. Therefore, blocked sampling can be efficiently performed with a flexible block size determined by the provided standard deviation.

%

\section*{Acknowledgment}
The author would like to thank Dr. Cong Ling (Imperial College London, UK), Prof. Guillaume Hanrot and Prof. Damien Stehl\'{e} (\'{E}cole Normale Sup\'{e}rieure de Lyon, France) for their helpful discussions and insightful suggestions.

\bibliographystyle{IEEEtran}
\bibliography{IEEEabrv,reference1}

\begin{thebibliography}{10}
\providecommand{\url}[1]{#1}
\csname url@samestyle\endcsname
\providecommand{\newblock}{\relax}
\providecommand{\bibinfo}[2]{#2}
\providecommand{\BIBentrySTDinterwordspacing}{\spaceskip=0pt\relax}
\providecommand{\BIBentryALTinterwordstretchfactor}{4}
\providecommand{\BIBentryALTinterwordspacing}{\spaceskip=\fontdimen2\font plus
\BIBentryALTinterwordstretchfactor\fontdimen3\font minus
  \fontdimen4\font\relax}
\providecommand{\BIBforeignlanguage}[2]{{%
\expandafter\ifx\csname l@#1\endcsname\relax
\typeout{** WARNING: IEEEtran.bst: No hyphenation pattern has been}%
\typeout{** loaded for the language `#1'. Using the pattern for}%
\typeout{** the default language instead.}%
\else
\language=\csname l@#1\endcsname
\fi
#2}}
\providecommand{\BIBdecl}{\relax}
\BIBdecl

\bibitem{Banaszczyk}
W.~Banaszczyk, ``New bounds in some transference theorems in the geometry of
  numbers,'' \emph{Math. Ann.}, vol. 296, pp. 625--635, 1993.

\bibitem{Forney_89}
G.~Forney and L.-F. Wei, ``Multidimensional constellations--{Part II}: Voronoi
  constellations,'' \emph{IEEE J. Sel. Areas Commun.}, vol.~7, no.~6, pp.
  941--958, Aug. 1989.

\bibitem{Kschischang_Pasupathy}
F.~R. Kschischang and S.~Pasupathy, ``Optimal nonuniform signaling for
  {G}aussian channels,'' \emph{{IEEE} Trans. Inform. Theory}, vol.~39, pp.
  913--929, May. 1993.

\bibitem{LiuLing2}
L.~Liu and C.~Ling, ``Polar codes and polar lattices for independent fading
  channels,'' \emph{IEEE Transactions on Communications}, vol.~64, no.~12, pp.
  4923--4935, Dec 2016.

\bibitem{LB_13}
C.~Ling and J.-C. Belfiore, ``Achieiving the {AWGN} channel capacity with
  lattice {Gaussian} coding,'' \emph{IEEE Trans. Inform. Theory}, vol.~60,
  no.~10, pp. 5918--5929, Oct. 2014.

\bibitem{LiuLing1}
L.~Liu, Y.~Yan, and C.~Ling, ``Achieving secrecy capacity of the {G}aussian
  wiretap channel with polar lattices,'' \emph{IEEE Transactions on Information
  Theory}, vol.~64, no.~3, pp. 1647--1665, March 2018.

\bibitem{LLBS_12}
C.~Ling, L.~Luzzi, J.-C. Belfiore, and D.~Stehl\'{e}, ``Semantically secure
  lattice codes for the {G}aussian wiretap channel,'' \emph{IEEE Trans. Inform.
  Theory}, vol.~60, no.~10, pp. 6399--6416, Oct. 2014.

\bibitem{7360779}
H.~Mirghasemi and J.~C. Belfiore, ``Lattice code design criterion for mimo
  wiretap channels,'' in \emph{Proc IEEE Information Theory Workshop (ITW)},
  Oct 2015, pp. 277--281.

\bibitem{7058433}
S.~Vatedka, N.~Kashyap, and A.~Thangaraj, ``Secure compute-and-forward in a
  bidirectional relay,'' \emph{IEEE Transactions on Information Theory},
  vol.~61, no.~5, pp. 2531--2556, May 2015.

\bibitem{ProbabilisticShapingOptical1}
T.~Fehenberger, D.~Lavery, R.~Maher, A.~Alvarado, P.~Bayvel, and N.~Hanik,
  ``Sensitivity gains by mismatched probabilistic shaping for optical
  communication systems,'' \emph{IEEE Photonics Technology Letters}, vol.~28,
  no.~7, pp. 786--789, Apr. 2016.

\bibitem{ProbabilisticShapingOptical2}
T.~Fehenberger, A.~Alvarado, G.~B\"ocherer, and N.~Hanik, ``On probabilistic
  shaping of quadrature amplitude modulation for the nonlinear fiber channel,''
  \emph{Journal of Lightwave Technology}, vol.~34, no.~21, pp. 5063--5073, July
  2016.

\bibitem{MicciancioGaussian}
D.~Micciancio and O.~Regev, ``Worst-case to average-case reductions based on
  {Gaussian} measures,'' in \emph{Proc. Ann. Symp. Found. Computer Science},
  Rome, Italy, Oct. 2004, pp. 372--381.

\bibitem{CHKP10}
D.~Cash, D.~Hofheinz, E.~Kiltz, and C.~Peikert, ``Bonsai trees, or how to
  delegate a lattice basis,'' in \emph{Proc. EUROCRYPT}, vol.~6, 2010, pp.
  523--552.

\bibitem{LwelatticeGaussian}
C.~P. V.~Lyubashevsky and O.~Regev, ``On ideal lattices and learning with
  errors over rings,'' in \emph{Proc. EUROCRYPT}, vol.~6, 2010, pp. 1--23.

\bibitem{GentrySW13}
C.~Gentry, A.~Sahai, and B.~Waters, ``Homomorphic encryption from learning with
  errors: Conceptually-simpler, asymptotically-faster, attribute-based,'' in
  \emph{CRYPTO}, Springer, Heidelberg, pp. 75-92, 2013.

\bibitem{latticeGaussianSignatures}
T.~Oder, T.~P\"{o}ppelmann, and T.~G\"{u}neysu, ``Beyond {ECDSA} and {RSA}:
  {L}attice-based digital signatures on constrained devices,'' in \emph{Proc.
  51st Annual Design Automation Conference}, 2014, pp. 1--6.

\bibitem{RegevSolvingtheShortestVectorProblem}
D.~Aggarwal, D.~Dadush, O.~Regev, and N.~Stephens-Davidowitz, ``Solving the
  shortest vector problem in $2^n$ time via discrete {Gaussian} sampling,''
  \emph{STOC}, 2015.

\bibitem{RegevSolvingtheClosestVectorProblem}
D.~Aggarwal, D.~Dadush, and N.~Stephens-Davidowitz, ``Solving the closest
  vector problem in $2^n$ time --- the discrete {Gaussian} strike again!''
  \emph{FOCS}, 2015.

\bibitem{Klein}
P.~Klein, ``Finding the closest lattice vector when it is unusually close,'' in
  \emph{ACM-SIAM Symp. Discr. Algorithms}, 2000, pp. 937--941.

\bibitem{ZhengWangTIT17}
Z.~Wang and C.~Ling, ``Lattice {G}aussian sampling by {M}arkov chain {M}onte
  {C}arlo: Bounded distance decoding and trapdoor sampling,'' \emph{Submitted
  to IEEE Transactions on Information Theory}, [Online]
  Available:http://arxiv.org/abs/1704.02673.

\bibitem{DerandomizedJ}
Z.~Wang, S.~Liu, and C.~Ling, ``Decoding by sampling - {Part} {II}:
  Derandomization and soft-output decoding,'' \emph{IEEE Trans. Commun.},
  vol.~61, no.~11, pp. 4630--4639, Nov. 2013.

\bibitem{CongRandom}
S.~Liu, C.~Ling, and D.~Stehl\'{e}, ``{Decoding by sampling: A randomized
  lattice algorithm for bounded distance decoding},'' \emph{IEEE Trans. Inform.
  Theory}, vol.~57, pp. 5933--5945, Sep. 2011.

\bibitem{Shaoshi1}
S.~Yang and L.~Hanzo, ``Fifty years of {MIMO} detection: The road to
  large-scale {MIMO}s,'' \emph{IEEE Communications Surveys Tutorials}, vol.~17,
  no.~4, pp. 1941--1988, Sep. 2015.

\bibitem{Luo1}
K.~Luo and A.~Manikas, ``Joint transmitter¨creceiver optimization in
  multitarget {MIMO} radar,'' \emph{IEEE Transactions on Signal Processing},
  vol.~65, no.~23, pp. 6292--6302, Dec 2017.

\bibitem{xia2}
H.~Cheng, Y.~Xia, Y.~Huang, L.~Yang, and D.~P. Mandic, ``A normalized complex
  {LMS} based blind {I}/{Q} imbalance compensator for {GFDM} receivers and its
  full second-order performance analysis,'' \emph{IEEE Trans. on Signal
  Process.}, vol.~66, no.~17, pp. 4701--4712, Sep. 2018.

\bibitem{Wuqihui1}
Q.~Wu, G.~Ding, J.~Wang, and Y.~Yao, ``Spatial-temporal opportunity detection
  for spectrum-heterogeneous cognitive radio networks: Two-dimensional
  sensing,'' \emph{IEEE Transactions on Wireless Communications}, vol.~12,
  no.~2, pp. 516--526, Feb. 2013.

\bibitem{Zhuang1}
J.~Zhuang, H.~Xiong, W.~Wang, and Z.~Chen, ``Application of manifold separation
  to parametric localization for incoherently distributed sources,'' \emph{IEEE
  Transactions on Signal Processing}, vol.~66, no.~11, pp. 2849--2860, June
  2018.

\bibitem{Xiangmin1}
M.~Xiang, B.~S. Dees, and D.~P. Mandic, ``Multiple-model adaptive estimation
  for 3-{D} and 4-{D} signals: A widely linear quaternion approach,''
  \emph{IEEE Transactions on Neural Networks and Learning Systems}, pp. 1--13,
  2018.

\bibitem{ZhengWangTIT15}
Z.~Wang and C.~Ling, ``On the geometric ergodicity of {M}etropolis-{H}astings
  algorithms for lattice {G}aussian sampling,'' \emph{IEEE Transactions on
  Information Theory}, vol.~64, no.~2, pp. 738--751, Feb. 2018.

\bibitem{ZhengWangMCMCLatticeGaussian}
Z.~Wang, C.~Ling, and G.~Hanrot, ``{Markov chain Monte Carlo} algorithms for
  lattice {Gaussian} sampling,'' in \emph{Proc. IEEE International Symposium on
  Information Theory (ISIT)}, Honolulu, USA, Jun. 2014, pp. 1489--1493.

\bibitem{ITW2017}
Z.~Wang and C.~Ling, ``On the geometric ergodicity of {G}ibbs algorithm for
  lattice {G}aussian sampling,'' in \emph{Proc. IEEE Information Theory
  Workshop (ITW)}, 2017, pp. 269--273.

\bibitem{Trapdoor}
C.~Gentry, C.~Peikert, and V.~Vaikuntanathan, ``Trapdoors for hard lattices and
  new cryptographic constructions,'' in \emph{Proc. 40th Ann. ACM Symp. Theory
  of Comput.}, Victoria, Canada, 2008, pp. 197--206.

\bibitem{LLLoriginal}
A.~K. Lenstra, H.~W. Lenstra, and L.~Lovasz, ``Factoring polynomials with
  rational coefficients,'' \emph{Math. Annalen}, vol. 261, pp. 515--534, 1982.

\bibitem{Shanxiang1}
S.~Lyu and C.~Ling, ``Boosted {KZ} and {LLL} algorithms,'' \emph{IEEE
  Transactions on Signal Processing}, vol.~65, no.~18, pp. 4784--4796, Sept
  2017.

\bibitem{mixingtimemarkovchain}
D.~A. Levin, Y.~Peres, and E.~L. Wilmer, \emph{Markov Chains and Mixing Time},
  American Mathematical Society, 2008.

\bibitem{RosenthalOPDAMCMC}
J.~S. Rosenthal, ``Optimal proposal distributions and adaptive {MCMC},''
  \emph{Handbook of Markov chain Monte Carlo: Methods and Applications.},
  Brooks, S.P., Gelman, A., Jones, G., Meng, X.-L. (eds.) Chapman and Hall/CRC
  Press, Florida, USA. 2010.

\bibitem{LukeTierney1}
L.~Tierney, ``Markov chains for exploring posterior distributions (with
  discussion),'' in \emph{Proc Ann. Stat.}, vol.~22, 1994, pp. 1701--1762.

\bibitem{Meynbook}
S.~P. Meyn and R.~L. Tweedie, \emph{Markov chains and stochastic
  stability}.\hskip 1em plus 0.5em minus 0.4em\relax UK, Cambridge University
  Press, 2009.

\bibitem{GibbsOriginal}
S.~Geman and D.~Geman, ``Stochastic relaxation, {G}ibbs distributions, and the
  {B}ayesian restoration of images,'' \emph{IEEE Transactions on Pattern
  Analysis and Machine Intelligence}, vol.~6, no.~6, pp. 721--741, 1984.

\bibitem{RobertsGeneralstatespace}
G.~O. Roberts, ``General state space {Markov} chains and {MCMC} algorithms,''
  \emph{Probability Surveys}, vol.~1, pp. 20--71, 2004.

\bibitem{SpectralGapL2}
I.~Kontoyannis and S.~P. Meyn, ``Geometric ergodicity and spectral gap of
  non-reversible real valued {M}arkov chains,'' in \emph{Proc. Probab. Theory
  and related Fields}, vol. 154, 2012, pp. 327--339.

\bibitem{LiuBook}
J.~S. Liu, \emph{Monte Carlo Strategies in Scientific Computing}, New York:
  Springer-Verlag, 2001.

\bibitem{Fill1991}
J.~A. Fill, ``Eigenvalue bounds on convergence to stationary for nonreversible
  markov chains, with application to the exclusion process.'' in \emph{Proc.
  Annals of Applied Probability}, vol.~1, 1991, pp. 62--87.

\bibitem{LiuCovarianceScans}
J.~S. Liu, W.~H. Wong, and A.~Kong, ``Covariance structure and convergence rate
  of the {Gibbs} sampler with various scans,'' \emph{J. Roy. Statist. Soc.
  Series B}, \textbf{57}(1): 157-169, 1995.

\bibitem{LiuFraction}
J.~S. Liu, ``Fraction of missing information and convergence rate of data
  augmentation,'' in \emph{Computationally Intensive Statistical Methods:
  Proceedings of the 26th symposium on the Interface}, North Carolina, 1994,
  pp. 490--497.

\bibitem{MetropolisOrignial}
N.~Metropolis, A.~W. Rosenbluth, M.~N. Rosenbluth, A.~H. Teller, and E.~Teller,
  ``{Equations of state calculations by fast computing machines},'' \emph{J.
  Chem. Phys.}, vol.~21, pp. 1087--1091, 1953.

\bibitem{Hastings1970}
W.~K. Hastings, ``{Monte Carlo} sampling methods using {Markov} chains and
  their applications,'' \emph{Biometrika}, vol.~57, pp. 97--109, 1970.

\bibitem{LiuPeskun}
J.~S. Liu, ``Peskun's theorem and a modified discrete-state {G}ibbs sampler,''
  \emph{Biometrika}, vol.~83, no.~3, pp. 681--682, 1996.

\bibitem{orderingmcmc}
A.~Mira, ``Ordering and improving the performance of {M}onte {C}arlo {M}arkov
  chains,'' \emph{Statistical Science}, vol.~16, no.~4, pp. 340--350, 2001.

\bibitem{ParallelTempering2}
C.~J. Geyer and E.~A. Thompson, ``Annealing {M}arkov chain {M}onte {C}arlo with
  applications to ancestral inference,'' \emph{J. Amer. Statist. Assoc.},
  vol.~90, pp. 909--920, 1995.

\bibitem{ParallelTempering1}
D.~J. Earl and M.~W. Deem, ``Parallel tempering: theory, applications, and new
  perspectives,'' \emph{Phys. Chem. Chem. Phys.}, vol.~7, pp. 3910--3916, 2005.

\bibitem{Regevlearning}
O.~Regev, ``On lattice, learning with errors, random linear codes, and
  cryptography,'' \emph{J. ACM}, vol.~56, no.~6, pp. 34:1--34:40, 2009.

\bibitem{BrakerskiLPRS13}
Z.~Brakerski, A.~Langlois, C.~Peikert, O.~Regev, and D.~Stehl{\'e}, ``Classical
  hardness of learning with errors,'' in \emph{STOC}, 2013, pp. 575--584.

\bibitem{HassibiMCMCnew}
B.~Hassibi, M.~Hansen, A.~Dimakis, H.~Alshamary, and W.~Xu, ``Optimized {Markov
  Chain Monte Carlo} for signal detection in {MIMO} systems: {An} analysis of
  the stationary distribution and mixing time,'' \emph{IEEE Transactions on
  Signal Processing}, vol.~62, no.~17, pp. 4436--4450, Sep. 2014.

\bibitem{MCMCKumarGibbsStalling}
A.~Kumar, S.~Chandrasekaran, A.~Chockalingam, and B.~S. Rajan, ``Near-optimal
  large-{MIMO} detection using randomized {MCMC} and randomized search
  algorithms,'' in \emph{Proc. IEEE International Conference on Communications
  (ICC)}, June. 2011.

\bibitem{xia1}
Y.~Xia and D.~P. Mandic, ``Augmented performance bounds on strictly linear and
  widely linear estimators with complex data,'' \emph{IEEE Trans. on Signal
  Process.}, vol.~66, no.~2, pp. 507--514, Jan 2018.

\bibitem{Babai}
L.~Babai, ``On {L}ov\'asz' lattice reduction and the nearest lattice point
  problem,'' \emph{Combinatorica}, vol.~6, no.~1, pp. 1--13, 1986.

\end{thebibliography}

\end{document}